\documentclass[10pt,prd,twocolumn]{revtex4-1}  
\usepackage{hyperref,graphicx,color,amsmath, amssymb}
\def \kpiv{k_*}
\def \phipiv{\phi_{*}}
\def \lnkring{\ln ({k}_{\rm ring}/\kpiv)}
\def \primsca{\mathcal{P}_s}  
\def \primten{\mathcal{P}_t} 
\def \amp#1{\left\vert{#1}\right\vert}

\def \halffigurewidth{0.45\textwidth}
\newcommand{\vk}{\mathbf{k}}
\newcommand{\vx}{\mathbf{x}}
\newcommand{\be}{\begin{equation}}
\newcommand{\ee}{\end{equation}}

\newcommand{\A}{{\delta \ln A_s}}

\begin{document}
\title{Constraining inflation with future galaxy redshift surveys}
\author{Zhiqi Huang$^1$, Licia Verde$^{2,3}$, Filippo Vernizzi$^1$}
\affiliation{${}^1$ CEA, Institut de Physique Th{\'e}orique, 91191 Gif-sur-Yvette c{\'e}dex, France\\ CNRS, URA 2306, F-91191 Gif-sur-Yvette, France}
\affiliation{${}^2$ Institute of Sciences of the Cosmos (ICCUB) 
 University of Barcelona, Marti i Franques 1,
Barcelona 08024, Spain}
\affiliation{${}^3$ICREA (Instituci\'o Catalana de Recerca i Estudis Avan\c{c}at)}
\date{\today}
\begin{abstract}
With future galaxy surveys, a huge number of Fourier modes of the distribution of the large scale structures in the Universe will become available. These modes are complementary to those of the CMB and can be used to set constraints on models of the early universe, such as inflation. 
Using a MCMC analysis, we compare the  power of the CMB with that  of the combination of CMB and galaxy survey data, to constrain the power spectrum of primordial fluctuations generated during inflation. We base our analysis on the Planck satellite and a spectroscopic redshift survey with  configuration parameters close to those of the Euclid mission as examples. We first consider models of slow-roll inflation, and show that the inclusion of large scale structure data  improves the constraints by nearly halving the error bars on the scalar spectral index and its running.  If we attempt to reconstruct the inflationary single-field potential, a similar conclusion can be reached on the parameters characterizing the potential.
We then study models with features in the power spectrum. In particular, we consider ringing features produced by a break in the potential and oscillations such as in axion monodromy. Adding large scale structures improves the constraints on features by more than  a factor of two. In axion monodromy we show that there are oscillations with small amplitude and frequency in momentum space that are undetected by CMB alone but can be measured  by including galaxy surveys in the analysis.
\end{abstract}

\maketitle

\section{Introduction}

Inflation \cite{Guth:1981, Guth/Pi:1982, Starobinsky:1982} is a successful paradigm to explain the observed Cosmic Microwave Background (CMB) anisotropies. Currently, the Planck satellite \footnote{http://www.rssd.esa.int/index.php?project=Planck} is taking data and one of its main goals will be to accurately measure the primordial power spectrum of scalar and tensor fluctuations, with the aim of constraining the inflationary parameters and, possibly, ruling out  models.
Indeed, the spectrum of primordial perturbations represents an important source of information on the inflationary phase: during inflation scalar and tensor fluctuations are produced with amplitudes and shapes  related to the dynamics of the fields driving inflation. 
For instance, slow-roll inflation predicts a firm deviation from a purely scale-invariant spectrum, i.e.~$n_s=1$. Indeed, by combining the WMAP CMB data with BAO distance measure \cite{Percival/etal:2010} and the Hubble constant $H_0$ measurement \cite{Riess/etal:2009}, the WMAP team reported a deviation of $n_s-1 = -0.037\pm 0.012$, a measurement excluding the purely scale-invariant spectrum by more than 3$\sigma$ \cite{Komatsu/etal:2011}. 

On the other hand, a series of new surveys  is being planned to accurately measure the large scale structure  of the Universe. The number of Fourier modes available in these surveys will be larger than those collected by the Planck satellite.  Furthermore, these modes will be on scales smaller than those  probed by the CMB (on scales where primary  anisotropies dominate) and sample the full three dimensional (3-D) structure of the density field. Even with the drawback that the late-time density field has evolved gravitationally and that non-linearities severely  limit  the amount of information that can be extracted from large-scale structure surveys, their large number of modes make them an invaluable observable.
 Hence, galaxy surveys are complementary probes to the CMB anisotropies and any attempt of constraining the initial conditions  should  take them into account. The literature on adding large-scale structure to CMB data to constrain inflationary model is extensive, e.g., 
 \cite{Peiris/etal:2003,Easther/Peiris:2006,Peiris/Easther:2008,Lesgourgues/Valkenburg:2007,Finelli/etal:2010} and references therein.

In this paper we perform a Markov Chain Monte Carlo (MCMC) analysis  of the constraining power on the primordial power spectrum gained by combining Planck data with those from a future spectroscopic  galaxy redshift survey with  Euclid-like \footnote{http://www.euclid-ec.org} characteristics.  This approach is more time-consuming and numerically intensive than the popular Fisher matrix approach to forecasts, but it is more precise and robust  and  solves many of the  drawbacks of Fisher-based forecasts.
As explained in Sec.~\ref{sec:nsnrun}, for Planck we use  2.5 years multiple channel mock data \cite{PlanckScienceTeam:2009} while for the galaxy survey we use    a survey with specifications described in Sec \ref{sec:LSS}  as an example of stage IV surveys, according to the classification of \cite{Albrecht/etal:2006}. Such specifications are  similar (although not identical)  to those of Euclid  \cite{Laureijs/etal:2009, Laureijs/etal:2011}.  Hereafter we refer to such a data set as LSS.

We only consider  single-field models of inflation where the inflaton field which drives inflation is a canonical scalar field characterized by some potential.
In the literature there are mainly three approaches to constraining inflation. The simplest is to assume that the inflaton field underwent a period of slow-roll during the phase where the observable primordial perturbations were generated, as reviewed in Sec.~\ref{sec:theory}. This allows  to connect {\em observations-based} parameters, which directly  characterize the primordial power spectrum, such as its amplitude, its tilt and the  running of the tilt, to {\em potential-based} parameters, written in terms of slow-roll quantities.
We follow this approach in Sec.~\ref{sec:results_ns}, where we show that LSS surveys improve the constraints on both the spectral tilt and its running by a factor of about $1.5$. 

A more sophisticated approach is to try to reconstruct the inflaton potential from the data. Many methods, reviewed in Sec.~\ref{sec:reconstruct}, have been proposed in the literature. All of them require some assumptions on the smoothness of the potential and their conclusions always depend on these priors. In our potential reconstruction we will assume that the inflaton potential remains smooth over a field range of the order of the Planck mass and, following Refs. \cite{Stewart:2002, Dodelson/Stewart:2002}, that also the first two slow-roll parameters -- those already constrained by the data -- remain  small. We use this method to forecast the power of reconstructing the inflaton potential up to its fourth derivative. For the first three derivatives we reach analogous conclusions to those drawn in the slow-roll case, although worsened by the addition of one extra parameter.

The third approach is to assume  a specific model -- given in terms of the inflaton potential -- characterized by a handful of parameters. In this approach one estimates the values and errors of these parameters from the data. We use this scheme in Sec.~\ref{sec:glitches} to study models generating features in the primordial spectrum.
Local signatures in the power spectrum can arise, for instance, when the inflaton potential has sharp features \cite{Starobinsky:1992,Chen/etal:2007, Chen/etal:2008}, when there is  a transition between different stages in the inflaton evolution  \cite{Contaldi/etal:2003,Cline/etal:2003}, when more than one field is present \cite{Polarski/Starobinsky:1992,Langlois/Vernizzi:2005}, from particle production during inflation \cite{Barnaby/etal:2009a, Barnaby/Huang:2009}, 
modulated preheating ~\cite{Chambers/Rajantie:2008,Bond/etal:2009}, 
or, more recently, in models motivated by monodromy in the extra dimensions \cite{Silverstein/Westphal:2008} (see also \cite{Bean/etal:2008}).
These features represent an important window on new physics because they are often related to UV scale phenomena inaccessible to Earth-based experiments. Furthermore, as single-field slow-roll inflation predicts a smooth power spectrum, it is important to estimate our power of falsifying this simplest model \cite{HuntSarkar2006,Covietal2006,MartinRingeval2004, Hamann/etal:2008, Pahud/etal:2009}.

In Sec.~\ref{sec:analytical} we discuss what are the main theoretical limitations to measure features using the CMB and the large-scale structure. For the large-scale structure these are the size of the volume of the survey, which limits the smallest measurable width of the features in momentum space, and the number of modes contained in the survey. The latter is limited by our understanding and control of the small non-linear scales, where a large number of modes is present. The CMB angular power spectrum is an integrated observable: the $C_\ell$ are  a 2-D quantity arising from a  convolution of the 3-D primordial power spectrum with the  radiation transfer function.
Small scales features get smoothed by this projection process and may remain unresolved \cite{Adshead/etal:2011b}.

Two typical examples of features in the primordial power spectrum are  ringing signatures from a step in the potential \cite{Starobinsky:1992} (Sec.~\ref{sec:ringing}) and superimposed oscillations from  axion monodromy \cite{Silverstein/Westphal:2008,McAllister/etal:2010, Flauger/etal:2010} (Sec.~\ref{sec:monodromy}). We discuss these two cases and find that adding large-scale structure to the CMB may be crucial in detecting features. In particular, in the case of oscillations we show that, as a function of the frequency, the likelihood is very different from a Gaussian. Thus, detecting the correct oscillation becomes more similar to the process of tuning a radio channel than to a Fisher matrix analysis. 
In this case we show that there are features remaining undetected even by variance-limited CMB data that can be measured with very high precision when LSS data are  added. In Sec.~\ref{sec:conclusion} we draw our conclusions. 

For the MCMC analysis employed in this paper we used a modified version of CosmoMC \cite{Lewis/Bridle:2002}. Contrarily to the original code, this modified version allows some of the parameters to have periodic boundary conditions. Moreover, for an accurate computation of the CMB angular power spectrum from a primordial spectrum with features, the standard publicly available CMB codes \cite{Lewis/etal:2000, Lesgourgues:2011, Blas/etal:2011, Doran:2005} are not sufficient. One of us (ZH) has developed a new numerical CMB integrator, built {\it ad hoc} to treat sharp features and frequent oscillations. The whole package is described  shortly in Appendix~\ref{sec:tech} and more extensively in Ref.~\cite{Huang:2012}.

\section{Forecasts formalism and observables \label{sec:nsnrun}}

In the following we  perform a MCMC calculation of the $68.3 \%$ and $95.4\%$ confidence level (CL) forecasts, using ``mock'' power spectra and the associated  predicted errors for the surveys considered.  For the CMB we use the expected signal to noise of the  Planck satellite \cite{PlanckScienceTeam:2009} while, as an example of upcoming galaxy surveys, we focus on a  stage IV  galaxy survey  (following the convention of Ref.~\cite{Albrecht/etal:2006}).  In particular, as ``straw-man"  survey we use  the specifications reported in Sec.~\ref{sec:LSS} which are  similar  to those of  Euclid  \cite{Laureijs/etal:2009,Laureijs/etal:2011}.
However, our findings will qualitatively  apply to this more general class of surveys,  which are likely to have comparable performances.
Furthermore, we only consider the spectroscopic component of the survey and only make use of the galaxy power spectrum. In particular, we will not directly use the weak gravitational lensing power spectrum, although weak lensing measurements will be useful in confirming some of the assumptions used here, such as for instance the  scale-independence of galaxy bias on linear scales. 

As fiducial cosmology for our analysis we use a flat $\Lambda$CDM model with $\Omega_ch^2=0.1128$, $\Omega_bh^2 = 0.022$, $h=0.72$, $\sigma_8 = 0.8$ and $\tau_{\rm re}=0.09$, where $\Omega_b$ and $\Omega_c$ are the energy fraction of baryons and cold dark matter, respectively, at redshift zero, $h$ is the Hubble constant in units of $100\,{\rm km}/(\text{s\,Mpc})$ and $\tau_{\rm re}$ is the reionization optical depth. 

\subsection{Forecasts for Planck}

For our forecast analysis we use  the expected signal to noise of Planck 2.5 years (5-sky surveys) of CMB  multiple  channel data. We use three channels for Planck mock data and we assume that the other channels are used for foreground removal and thus do not provide cosmological information. We list the instrument characteristics of the  channels used in our analysis in Table~\ref{tbl:PlanckSpec}. We take the detector sensitivities and the values of the full width half maxima from the Planck ``Blue Book'' \cite{PlanckScienceTeam:2009}. 
\begin{table}
\caption{Planck  Instrument Characteristics for 30 months of integration.\label{tbl:PlanckSpec}}
\centering
\begin{tabular}{llll}
\hline
\hline
Channel Frequency (GHz) & 70 &100 & 143 \\
\hline
Resolution (arcmin) & 14 & 10 & 7.1 \\
Sensitivity  - intensity ($\mu K$)&  8.8 & 4.7 & 4.1 \\
Sensitivity - polarization ($\mu K$)& 12.5 & 7.5  & 7.8  \\
\hline
\end{tabular}
\end{table}

Given a likelihood function ${\cal L}$, we define $\chi^2 \equiv -2 \ln {\cal L}$.
For a nearly full-sky CMB experiment (here we use an observed fraction of sky $f_{\rm sky}=0.75$) $\chi^2$ can be approximated by \cite{Verde/etal:2006,Baumann/etal:2009}
\begin{equation}
\begin{split}
\label{chisq_CMB}
\chi^2 =& \sum_{\ell=\ell_{\rm min}}^{\ell_{\rm max}} (2\ell+1)f_{\rm sky}\, \left[ -3 + \frac{\hat{{\cal C}}_\ell^{BB}}{{\cal C}_\ell^{BB}}  + \ln\left(\frac{{\cal C}_\ell^{BB}}{\hat{{\cal C}}_\ell^{BB}}\right) \right.  \\
& \left.  + \frac{\hat{{\cal C}}_\ell^{TT}{\cal C}_\ell^{EE} + \hat{{\cal C}}_\ell^{EE}{\cal C}_\ell^{TT} - 2\hat{{\cal C}}_\ell^{TE}{\cal C}_\ell^{TE}}{{\cal C}_\ell^{TT}{\cal C}_\ell^{EE}-({\cal C}_\ell^{TE})^2} \right. \\ 
&+ \left. \ln{\left(\frac{{\cal C}_\ell^{TT}{\cal C}_\ell^{EE}-({\cal C}_\ell^{TE})^2}{\hat{{\cal C}}_\ell^{TT}\hat{{\cal C}}_\ell^{EE}-(\hat{{\cal C}}_\ell^{TE})^2}\right)}\right] \ ,  
\end{split}
\end{equation}
where we assume $l_{\rm min}=3$ (given the sky cut the error on $\ell=2$ is large)  and $l_{\rm max}=2500$. In this formula, ${\cal C}^{XY}_\ell$  are the model-dependent theoretical angular power spectra for the temperature, $E$ and $B$ polarizations and their cross-correlations, with $X,Y=\{T,E,B \}$. They are given by ${\cal C}^{XY}_\ell= C^{XY}_\ell + N_\ell^{XY}$, where the $C_\ell$ are calculated using the publicly available code CAMB \cite{Lewis/etal:2000} and the $N_\ell$ are noise spectra, which we compute assuming Gaussian beams. The estimators of the measured angular power spectra, $\hat{{\cal C}}^{XY}_\ell$, include the contribution from the noise.

We use the model introduced in \cite{Verde/etal:2006} (and later updated in \cite{Baumann/etal:2009}) to propagate the effect of polarization foreground residuals into the estimated uncertainties on the cosmological parameters. For simplicity, in our simulation we consider only the dominant components in the frequency bands that we are using, i.e., the synchrotron and dust signals.  We assume that foreground subtraction can be done correctly down to a level of 5\%.

\subsection{Forecasts for LSS surveys}
\label{sec:LSS}

We now turn to the LSS survey forecasts. 

We model the galaxy power spectrum in redshift space as (e.g., \cite{Kaiser:1987, Peacock:1992, Peacock/Dodds:1994})
\begin{equation}
P_g(k,\mu; z) = \left(b+ f \mu^2\right)^2 D^2(z)P_{\rm m}(k) \exp\left(-k^2\mu^2\sigma_r^2 \right), \label{eq:Pg}
\end{equation}
where $\mu$ is the cosine of the angle between the wavenumber $\mathbf{k}$ and the line of sight, $D(z)$ is the linear growth factor, $f \equiv {d\ln D}/{d\ln a}$ is the linear growth rate, $P_{\rm m}(k)$ is the matter power spectrum today (at redshift $z=0$) and $\sigma_r$ parameterizes the effect of small scales velocity dispersion and redshift errors as explained below.

The term $f \mu^2$  accounts for the redshift distortions due to the large-scale peculiar velocity field \cite{Kaiser:1987}, which is correlated with the matter density field. 
The exponential factor on the right-hand side  accounts for the radial smearing due to the redshift distortions that are uncorrelated with the large-scale structure. In particular, we consider two contributions. The first is due to the redshift uncertainty of the spectroscopic galaxy samples which is estimated to be $\sigma_z=0.001(1+z)$  \cite{Laureijs/etal:2009}. The second  comes from the Doppler shift due to the virialized motion of galaxies within clusters, which  typically has  a pairwise velocity dispersion $\sigma_g$ of the order of few hundred $\text{km/s}$. This can be parameterized as $\frac{\sigma_g}{\sqrt{2}} (1+z)$ \cite{Peacock/Dodds:1994}.
Taking the qudratic mean of the two contributions,
this turns into a comoving distance dispersion $\sigma_r$ given by  
\begin{equation}
\sigma_r^2 = \frac{(1+z)^2}{H^2(z)}  \left(10^{-6} + {\sigma_g^2}/{2} \right)\;, \label{eq:sigma2}
\end{equation}
where  $H$ is the Hubble parameter as a function of the redshift. Note that the two quadratic velocity dispersions in the parenthesis  are  of the same magnitude.
Practically, neither the redshift measurement nor the virialized motion of galaxies can be precisely modeled. In particular, the radial smearing due to peculiar velocity is not necessarily close to Gaussian. Thus, eq.~(\ref{eq:Pg}) should not be used for wavenumbers $k>\frac{H(z)}{\sigma_g (1+z)}$, where the radial smearing effect is important.  
This  consideration strictly applies only to the radial direction. However,  gravitational non-linearities affect equally the radial as well as the angular direction. By imposing this cut at all $\mu$  we make sure that we exclude from our analysis non-linear scales. 
We thus introduce a UV cutoff $k_{\max}$ as the smallest value between $\frac{H}{\sigma_g(1+z)}$ and $\frac{\pi}{2R}$, where $R$ is chosen such that the r.m.s.~linear density fluctuation of the matter field in a sphere with radius $R$ is $0.5$. The resulting maximum wave numbers  $k_{\rm max}$ are reported in Table~\ref{tbl:Euclid_LSS_zbins} for each redshift bin defined below. Note that these UV cutoffs are relatively conservative, even comparing to the current values used for the SDSS catalog at redshift $\sim 0.35$ \cite{Reid/etal:2010}.

\begin{table*}
\begin{center}
\caption{Redshift bins used in the analysis. With  current specifications, the Euclid mission  will not  include the first redshift bin; other ground-based surveys will, albeit with smaller sky coverage.  \label{tbl:Euclid_LSS_zbins}}
\begin{tabular}{cccccc}
\hline
\hline
\  $\bar{z}$ \ & \ $\Delta z$ \ & \ $\bar{n}_{\rm obs}$ ($h^3\,{\rm Mpc}^{-3}$) \ & \ $k_{\rm min}$ ($h\,{\rm Mpc}^{-1}$) \ & \ $k_{\rm max}$ ($h\,{\rm Mpc}^{-1}$) \ & \ bias $b$ \ \\
\hline
  0.6 & 0.2 & $3.56\times10^{-3}$ & 0.0033 & 0.15  &  1.053\\
  0.8 & 0.2 & $2.42\times10^{-3}$ & 0.0029 & 0.17 &  1.125 \\
  1.0 & 0.2 & $1.81\times10^{-3}$ & 0.0027 & 0.20 &  1.126 \\
  1.2 & 0.2 & $1.44\times10^{-3}$ & 0.0026 & 0.21 &  1.243 \\
  1.4 & 0.2 & $0.99\times10^{-3}$ & 0.0025 & 0.22 &  1.292 \\
  1.6 & 0.2 & $0.55\times10^{-3}$ & 0.0024 & 0.22 &  1.497 \\
  1.8 & 0.2 & $0.29\times10^{-3}$ & 0.0024 & 0.23 &  1.491 \\
  2.0 & 0.2 & $0.15\times10^{-3}$ & 0.0024 & 0.23 &  1.568 \\
\hline
\end{tabular}
\end{center}
\end{table*}

The survey volume is split into 8 redshift bins. The redshift ranges and expected numbers of observed galaxies per unit volume are given in Table~\ref{tbl:Euclid_LSS_zbins}. For the observed sky we use  $20'000$ sq.~degrees (Euclid is expected to only cover $15'000$ sq.~degrees, so this represents a departure from the baseline for Euclid).  With current specifications the Euclid mission  will not include the first redshift bin; other --ground based-- surveys  will, albeit with smaller sky coverage: $\sim 10'000$ sq.~degrees. However, the difference in  total survey volume between assuming $10'000$ sq.~degrees or $20'000$ sq.~degrees in the first redshift bin is only few percent and this has virtually no impact on our results.  The number density of galaxies that can be used is $\bar{n}=\varepsilon \cdot \bar{n}_{\rm obs}$, where $\varepsilon$ is the fraction of galaxies with measured redshift. 
Due to the high accuracy of the spectroscopic redshift and the width of the bins, we ignore the bin-to-bin correlations and write $\chi^2$ as 
\begin{equation}
  \chi^2 = \sum_{k,\mu,z\ \rm bins} \left(\frac{P_{g, \rm model} - P_{g, \rm fiducial}}{\Delta P_{g,\rm fiducial}}\right)^2\ .
\end{equation}
As on large scales the matter density field has, to a very good approximation, Gaussian statistics and uncorrelated Fourier modes, the band-power uncertainty is given by \cite{Tegmark/etal:1998} \footnote{We find that using the linear power spectrum approximation underestimates the error on $P(k)$ by less than 10--15\% for the scales and redshifts considered \cite{dePutter/etal:2011}}
\begin{equation}
\Delta  P_g = \left[ \frac{2 (2\pi)^3}{(2\pi k^2dk d\mu) (4\pi r^2f_{\rm sky} dr)}\right]^{1/2}\left(P_g+\frac{1}{\bar{n}}\right), \label{eq:dPg}
\end{equation}
where  $r$ is the comoving distance given, for a flat FRW universe, by $r(z)=\int_0^z cdz'/H(z')$. The second term in the parenthesis is due to shot noise, under the assumption that the positions of the observed galaxies are generated by a random Poisson point process. In practice $\bar{n}$ is not known a priori and is calibrated by galaxies themselves. The imperfect knowledge of $\bar{n}$ can bias $P_g$ on the scale of the survey \cite{Tegmark/etal:1998}. This is taken into account by using an IR cutoff $k_{\min}\sim {\rm Gpc}^{-1}$, chosen such that $k_{\min} ( z) =  2\pi/V^{1/3}( z)$, where $V( z)$ is the comoving volume of the redshift slice $\bar z -\Delta z/2 \le z \le \bar z +\Delta z/2$.  If not otherwise specified, in each redshift bin we use 30 $k$-bins uniformly in $\ln k$ and 20 uniform $\mu$-bins. For models with features in the primordial power spectrum, we use a different binning scheme in $k$, as explained in Section~\ref{sec:glitches}. For each redshift bin the value of $k_{\min}$ is reported in Table~\ref{tbl:Euclid_LSS_zbins}.

We consider a {\em pessimistic} and an {\em optimistic} case. For the pessimistic case we take $\varepsilon = 0.35$. 
For the fiducial value of the bias, in each redshift bin we take the one reported in the last column of Table~\ref{tbl:Euclid_LSS_zbins}.
We assume that $\sigma_g$ is redshift dependent and choose  $\sigma_g=400\,{\rm km/s}$ as the fiducial value in each redshift bin \footnote{Note that the authors of \cite{Giannatonio/etal:2011} used $\sigma_g = \sigma_{g,0} \sqrt{1+z}$, with $\sigma_{g,0}=400\,{\rm km/s}$, instead of a redshift-independent value for $\sigma_g$. Our choice differs from theirs only slightly at low redshift. This makes our analysis  more conservative.}. Then, we marginalize over $b$ and $\sigma_g$ in the 8 redshift bins, for a total of 16 nuisance parameters. For the optimistic case we take $\varepsilon=0.5$. For the fiducial value of the bias we use  $b=\sqrt{1+z}$, which is about 10-30\% larger than the one used in the pessimistic case. In this case we assume that $\sigma_g$ is redshift independent, so that there are only 9 nuisance parameters, i.e.~$b$ in 8 redshift bins and $\sigma_g$.

Gaussian statistics for the matter density field and a scale-indepedent bias in each redshift bin should be good approximations. On the other hand, we have discarded all information beyond the linear scales and marginalized over the bias. In reality, weak lensing could allow  us to get a good prior knowledge on the galaxy bias. Moreover, it is likely that we can obtain more information by modeling the mildly non-linear regime, by taking into account the bin-to-bin cross correlations (tomography) and using the three-point correlations. These topics are beyond the scope of this paper. In conclusion, the reader should bear in mind that the forecast made here is ``clean'' (linear-scale only) and hence ``conservative"\footnote{With the caveat that we do not include systematic errors, which may  in a practical application degrade the estimates}.

\section{Constraining smooth inflationary potentials} \label{sec:smooth}

\subsection{Primordial perturbations from inflation \label{sec:theory}}  

Let us start by introducing the standard parameterization of the primordial perturbations after inflation {(see for instance \cite{Liddle/Lyth:2000})}. In a perturbed flat FRW universe, by choosing a gauge where there are no fluctuations in the inflaton field we can write the spatial part of the metric as
\begin{equation}
g_{ij} = a^2(t)  \left[ \left( 1+ 2 \zeta \right) \delta_{ij} + h_{ij} \right] \;, \quad h_{ii} =0 \;, \quad \partial_i h_{ij}=0\;, 
\label{metric}
\end{equation}
where $a$ is the scale factor, $\zeta(t,\vx)$ is a scalar perturbation which has the property to be conserved on super-horizon scales for adiabatic perturbations, and $h_{ij}(t,\vx)$ is a spin-2 quantity charactering the tensor modes of the metric \footnote{The scalar quantity in eq.~(\ref{metric}) is sometimes called $\cal R$ and it is defined by ${\cal R} = - \Psi + H \delta u$, where $\Psi$ is the scalar curvature and $\delta u$ is the perturbed velocity potential, and $\zeta$ may denote the curvature perturbation on uniform density hypersurfaces, $\zeta = - \Psi - H \delta \rho/\dot \rho$, where $\rho$ is the energy density. On super-Hubble scales these two quantities are the same.}.

The power spectrum of primordial scalar perturbations, $P_\zeta(k)$, is defined by 
\begin{equation}
\langle \zeta_{\mathbf{k}} \; \zeta_{{\mathbf{k}}'} \rangle = (2 \pi)^3 \delta_{D}({\mathbf{k}}+{\mathbf{k}}') P_{\zeta}(k)\;,
\end{equation}
where $\langle \ldots \rangle$ denotes {an} ensemble average. It is useful to define a dimensionless spectrum for scalar fluctuations, 
\be
\primsca (k) \equiv \frac{k^3}{2\pi^2} P_\zeta (k)\;.
\ee
The deviation from scale-invariance of the scalar spectrum is characterized by the spectral index $n_s$, defined by
\begin{equation}
\label{eq:index_ns}
n_s \equiv 1 + \frac{d  \ln \primsca}{d \ln k}\;, 
\end{equation}
where $n_s=1$ denotes a purely scale-invariant spectrum. We also define the running of the spectral index ${\alpha_s}$ as
\begin{equation}
\label{eq:running}
{\alpha_s} \equiv \frac{d n_s}{d \ln k}\;.
\end{equation}
These quantities are defined at a particular pivot scale, which for our analysis we chose to be $\kpiv  \equiv 0.05 \, {\rm Mpc}^{-1}$. Thus, with these definitions the dimensionless power spectrum can be written as
\begin{equation}
{\cal P}_s(k) = A_s  \left(\frac{k}{\kpiv} \right)^{n_s(\kpiv ) -1 + \frac12 {\alpha_s}(\kpiv ) \ln (k/\kpiv ) }\;,
\label{eq:shape}
\end{equation}
where $A_s$ is a normalization parameterizing the amplitude of the fluctuations.

The tensor quantity $h_{ij}$ contains two modes of different polarization, $h_+$ and $h_\times$, each with power spectrum given by 
\begin{equation}
\langle h_{\mathbf{k}} h_{{\mathbf{k}}'} \rangle = (2 \pi)^3 \delta({\mathbf{k}}+{\mathbf{k}}') P_h(k)\;.
\end{equation}
Defining the dimensionless power spectrum of tensor fluctuations as $\primten (k) \equiv 2 \frac{k^3}{2\pi^2} P_h (k)\;$ (the factor of $2$ comes from the two polarizations), the ratio of tensor-to-scalar fluctuations is given by 
\begin{equation}
\label{eq:r_ts}
r\equiv \frac{\primten 
}{\primsca 
}\;. 
\end{equation}

During inflation the fractional change of the Hubble rate per $e$-fold is small, $\epsilon \equiv - \dot H/H^2 \ll 1$, where $H \equiv d \ln a/ dt$ is the Hubble rate and a dot denotes derivative with respect to the cosmic time.
Thus, the above observables take a simple form at leading order in $\epsilon$. The spectrum of fluctuations is given by (see \cite{Stewart/Lyth:1993} for an expression beyond leading order in $\epsilon$)
\begin{equation}
\primsca(k) = \left.\frac{1}{8 \pi^2 \epsilon}\frac{H^2}{M_p^2} \right|_{k=aH}\;, \label{eq:Psapprox}
\end{equation}
where $M_p \equiv (8 \pi G)^{-1/2}$ is the reduced Planck mass and the right-hand side is evaluated when the comoving scale $k$ exits the Hubble radius, while the spectral index \eqref{eq:index_ns} is given, using eqs.~\eqref{eq:index_ns} and \eqref{eq:Psapprox}, by $n_s = 1 -2 \epsilon - \dot \epsilon /(H\epsilon)|_{k=aH}$. The relative variation of $\epsilon$ in a Hubble time is typically small, so that the power spectrum is close to scale invariance.
The spectrum of tensor fluctuations is given by
\begin{equation}
\primten (k) = \left.\frac{2}{\pi^2 }\frac{H^2}{M_p^2} \right|_{k=aH}\;, \label{eq:Ptapprox}
\end{equation}
which shows that the ratio of tensor-to-scalar fluctuations in eq.~\eqref{eq:r_ts} is simply related to the first slow-roll parameter by $r =  16 \epsilon $.

In slow-roll inflation, during the adiabatic evolution one can related $\epsilon$ and $\dot \epsilon$ to the potential-based parameters $\epsilon_V$ and $\eta_V$, defined as 
\begin{equation}
\epsilon_V\equiv \frac{M_p^2}{2} \left( \frac{V'}{V} \right)^2 \; , \label{eq:epsetavdef} \qquad \eta_V \equiv M_p^2 \frac{V''}{V} \;, 
\end{equation}
where a prime denote the derivative with respect to the inflaton field. In terms of these parameters, the scalar spectral index reads, at leading order in slow-roll,
\begin{equation}
n_s = 1 - 6 \epsilon_V + 2 \eta_V \;. \label{eq:nsapprox}
\end{equation}
Another observational parameter {is the running of the spectral index introduced in eq.~\eqref{eq:running}}, which at leading order in slow-roll can be rewritten as 
\begin{equation}
{\alpha_s} = 16\epsilon \eta_V - 24 \epsilon_V^2- 2\xi_V \;, \label{eq:alphasapprox}
\end{equation}
where {$\xi_V$ is the third slow-roll parameter, defined as}
\begin{equation}
\xi_V \equiv M_p^4 \frac{V' V''' }{V^2} \; . \label{eq:xivdef}
\end{equation}
The running is thus second-order in slow-roll.


\subsection{Scalar spectral index and running in slow-roll models \label{sec:results_ns}}  

Let us compute the forecast on the spectral index $n_s$ and the scalar running $\alpha_s$ from slow-roll inflation. As a fiducial model we consider chaotic inflation based on the quadratic inflaton potential $V = \frac12 m^2 \phi^2$ \cite{Linde:1983}. In this case eq.~\eqref{eq:epsetavdef} gives $\epsilon_V=\eta_V = 2 M_p^2 /\phi^2 = 1/(2 N)$ (while from eq.~\eqref{eq:xivdef} $\xi_V =0$) and at leading order in slow-roll one finds, using eqs.~\eqref{eq:r_ts}, \eqref{eq:nsapprox} and \eqref{eq:alphasapprox},
\begin{equation}
n_s = 1 - 2/N\;, \quad r = 8/N\;, \quad {\alpha_s} = -2 /N^2\;,
\end{equation}
where $N$ is number of $e$-folds from Hubble crossing to the end of inflation.
Choosing $N=62.5$ this yields $n_s= 0.968$, $r=0.128$ and ${\alpha_s} =0$ as our fiducial model. This choice is fully consistent with a joint analysis of the latest cosmological data \cite{Komatsu/etal:2011, Reid/etal:2010, Huang/etal:2011, Castro/etal:2009}.

\begin{table*}
\caption{Cosmological Parameters \label{tbl:cosmoparams}}
\centering
\begin{tabular}{ccccc}
\hline
\hline
\  & \ Planck \ & \ Planck + LSS opt. \ & \ Planck + LSS pess. \ &\ Planck + LSS pess. Fisher \ \\
\hline
 $ \Omega_bh^2 $ & $0.02201^{+0.00012}_{-0.00012}$ & $0.02200^{+0.00008}_{-0.00008}$ & $0.02200^{+0.00008}_{-0.00008}$ & $0.02200^{+0.00008}_{-0.00008}$ \\
 $ \Omega_ch^2 $ & $0.1127^{+0.0010}_{-0.0009}$ & $0.11280^{+0.00021}_{-0.00023}$ & $0.11281^{+0.00026}_{-0.00026}$ & $0.1128^{+0.00025}_{-0.00025} $ \\
 $ \theta $ & $1.0460^{+0.0002}_{-0.0002}$ & $1.0460^{+0.0002}_{-0.0002}$ & $1.0460^{+0.0002}_{-0.0002}$ & $1.0460^{+0.0002}_{-0.0002}$\\
 $ \tau_{\rm re} $ & $0.0899^{+0.0041}_{-0.0038}$ & $0.0900^{+0.0030}_{-0.0030}$ & $0.0900^{+0.0030}_{-0.0030}$ &   $0.0900^{+0.0023}_{-0.0023}$ \\
 $ n_s $ & $0.9682^{+0.0030}_{-0.0029}$ & $0.9681^{+0.0017}_{-0.0017}$ & $0.968^{+0.0020}_{-0.0020}$ &  $0.968^{+0.0018}_{-0.0018}$ \\
 $ \alpha_s $ & $-0.0005^{+0.0050}_{-0.0050}$ & $-0.000^{+0.0030}_{-0.0030}$ & $-0.0001^{+0.0032}_{-0.0034}$ & $0.0000^{+0.0028}_{-0.0028}$ \\
 $ \ln (10^{10}A_s) $ & $3.019^{+0.009}_{-0.008}$ & $3.019^{+0.006}_{-0.006}$ & $3.019^{+0.006}_{-0.006}$ & $3.019^{+0.005}_{-0.005}$ \\
 $ r $ & $0.129^{+0.020}_{-0.020}$ & $0.128^{+0.020}_{-0.020}$ & $0.129^{+0.020}_{-0.020}$ & $0.128^{+0.018}_{-0.018}$ \\
 $ \Omega_m $ & $0.2595^{+0.0053}_{-0.0047}$ & $0.2599^{+0.0009}_{-0.0009}$ & $0.2600^{+0.0011}_{-0.0011}$ & - \\
 $ \sigma_8 $ & $0.7990^{+0.0046}_{-0.0046}$ & $0.7992^{+0.0026}_{-0.0026}$ & $0.7992^{+0.0027}_{-0.0028}$ & - \\
 $ H_0 $ & $72.05^{+0.45}_{-0.47}$ & $72.01^{+0.10}_{-0.09}$ & $72.01^{+0.11}_{-0.11}$ & - \\
\hline
\end{tabular}
\end{table*}
We consider the forecast constraints on eight cosmological parameters, i.e.~$\Omega_bh^2$, $\Omega_ch^2$, $\theta$, $\tau_{\rm re}$, $\ln A_s$, $n_s$, ${\alpha_s}$, and $r$. Here $\theta$ is the angle extended by sound horizon on the last scattering surface, rescaled by a factor $100$.  The nuisance parameters are marginalized over in the final result. The marginalized 68.3\% confidence level constraints on cosmological parameters for Planck forecast only, Planck and LSS optimistic forecast, and Planck and LSS pessimistic forecast are  listed in the second, third, and fourth columns of Table \ref{tbl:cosmoparams}, respectively. For comparison we list the error bars obtained from a Fisher matrix analysis in the fifth column.

\begin{figure}
\centering
\includegraphics[width=0.45\textwidth]{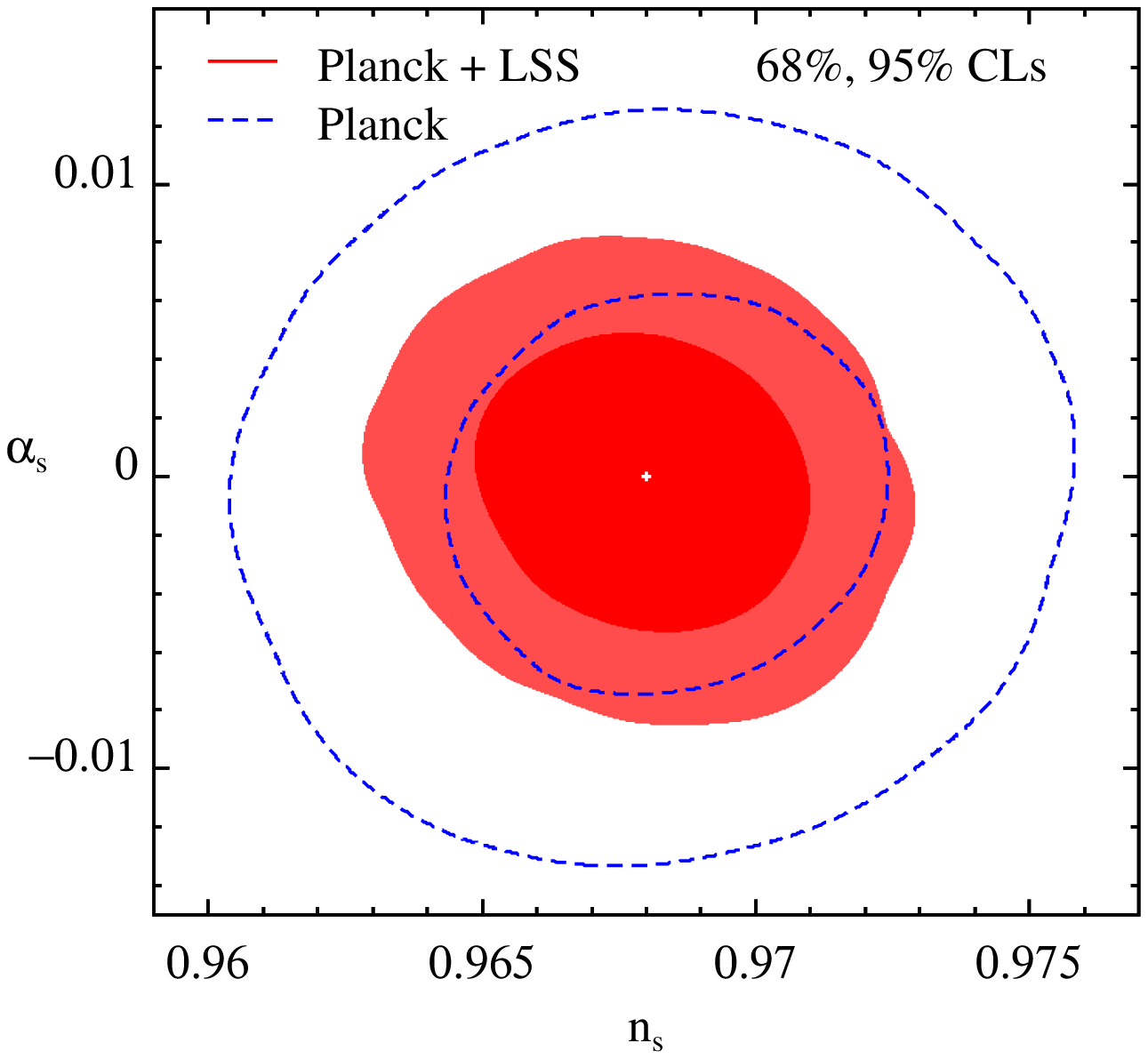}
\includegraphics[width=0.45\textwidth]{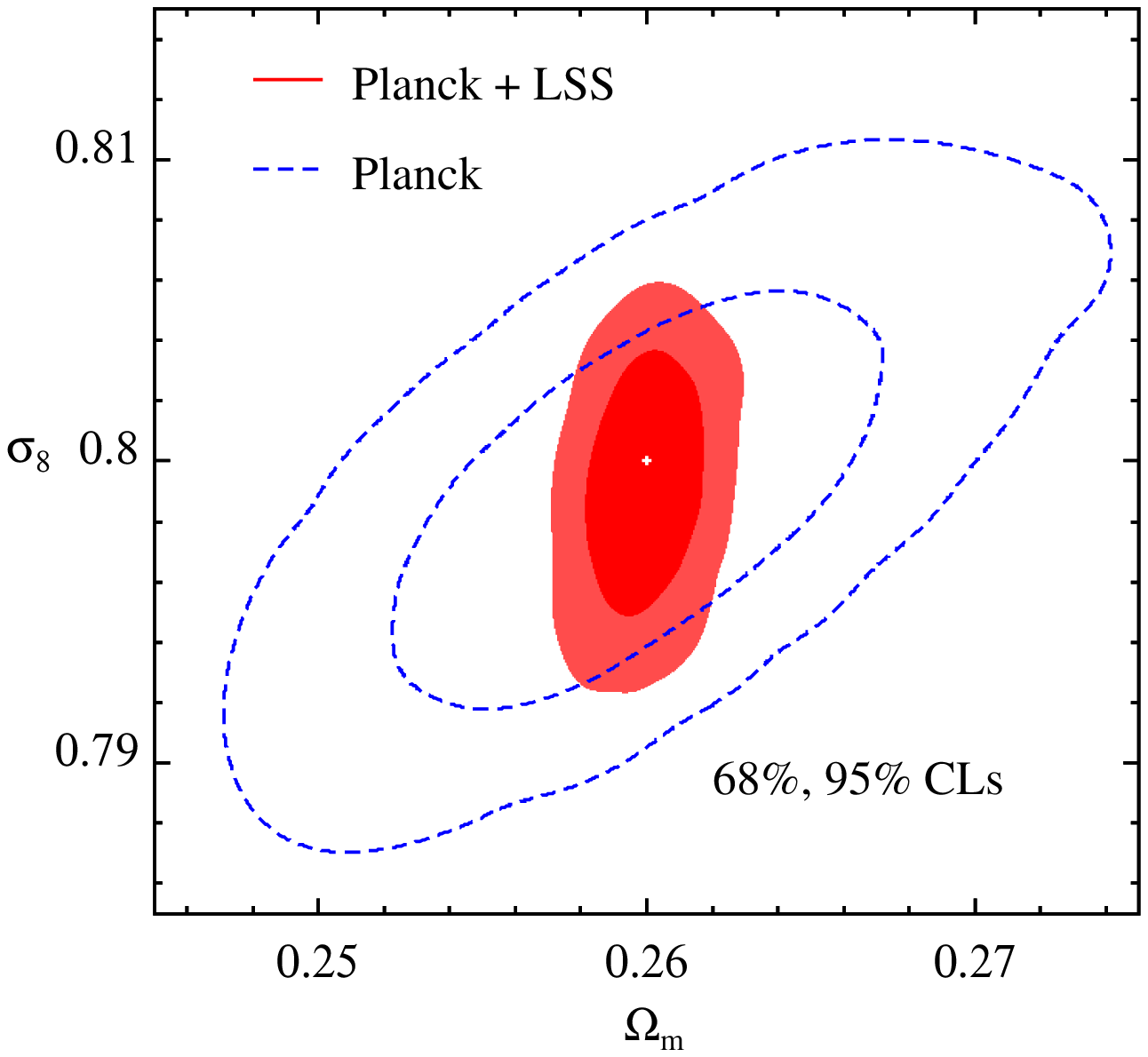}
\caption{Marginalized forecasted posterior contours (68.3\% and 95.4\% CL) on the $n_s$-$\alpha_s$ plane (top panel) and $\Omega_{\rm m}$-$\sigma_8$ plane (bottom panel) obtained for Planck alone (blue-dashed lines) and combining Planck with a pessimistic LSS survey (red-filled contours). The white dots correspond to the fiducial model. \label{fig:2D_Planck_Euclid}}
\end{figure}

The forecasted constraints on the plane $n_s$-${\alpha_s}$ are shown on the top panel of Fig.~\ref{fig:2D_Planck_Euclid}.
Even in the pessimistic case, LSS surveys  can improve the figure-of-merit for $\{n_s,{\alpha_s}\}$ (defined as ${\rm FOM} \equiv 1/\sqrt{\det{\rm Cov}(n_s,\alpha_s)}$ where Cov denotes the covariance matrix of the two parameters \cite{Wang:2008}) by a factor of $2.2$. Because the bias is unknown, large-scale structure data do not directly measure $A_s$ or $\sigma_8$. However, our straw-man LSS survey  can measure $\Omega_m$ to a much better accuracy, which can break the degeneracy between $\Omega_m$ and $\sigma_8$ that one typically finds using CMB data alone. This is shown in the bottom panel of Fig.~\ref{fig:2D_Planck_Euclid}. We have checked that for Planck data $r$ is almost orthogonal to $n_s$ and ${\alpha_s}$; hence, our result is not sensitive to the fiducial value of $r$. 

\subsection{Potential reconstruction \label{sec:reconstruct}}

Typically, slow-roll models predict a very small running of the scalar spectral index,  $\amp{\alpha_s}\lesssim 10^{-3}$. For these models the parametrization~(\ref{eq:shape}) is good enough. However, this prediction is based on the {\it assumption} that the inflaton potential is ``globally simple'', i.e.~that it does not abruptly change during the 60 $e$-folds of inflationary expansion. This assumption cannot { always be} tested, because the window of observed cosmological perturbations that ranges from the current Hubble radius to the scale where perturbations become non-linear is generated in about $ 10$ $e$-folds of inflation. Perturbations on smaller scales, generated in the remaining $\sim 50$ $e$-folds of inflation, are difficult to study because of the complicated non-linear dynamics in the late-time universe. As a consequence, only a small piece of the potential is directly observable with the CMB and the large-scale structure. 

One way to constrain potentials independently of the slow-roll assumption is to compute $V(\phi)$ for a given model and constrain the parameters of this model. We will follow this approach in the next section. In this section we consider a complementary procedure, called potential reconstruction, which consists in studying only the part of the potential responsible for the observable window and relaxing the assumption of slow-roll outside this window. The remaining part of the potential can be of any shape, as long as it gives the right number of $e$-folds of expansion.

Many methods have been proposed to reconstruct the observable piece of $V(\phi)$ \cite{Kinney:2002, Adshead/etal:2011, Mortonson/etal:2011, Adshead/Easther:2008, Agarwal/Bean:2009,Nicholson/etal:2010, Powell/Kinney:2007,Mukherjee/Wang:2003, Mukherjee/Wang:2005, Leach:2006, Leach/etal:2002, Leach/Liddle:2003, Peiris/Verde:2010, Peiris/Easther:2006, Peiris/Easther:2006a, Peiris/etal:2003, Easther/Peiris:2006, Easther/Kinney:2003, Verde/Peiris:2008, Nicholson/Contaldi:2009, Nagata/Yokoyama:2008, Tocchini-Valentini/etal:2005, Kogo/etal:2004, Shafieloo/Souradeep:2004, Shafieloo/Souradeep:2008, Bridges/etal:2006, Bridges/etal:2007, Spergel/etal:2007, Hannestad:2004, Bridle/etal:2003, Sealfon/etal:2005,Malquarti/etal:2004}. 
All of them require some assumption {regarding} the smoothness of the potential in the observable window, without which one could introduce arbitrarily many mass scales to build arbitrary shapes of the potential. The assumptions most used are: cutting at finite order some expansion in slow-roll parameters \cite{Kinney:2002, Peiris/etal:2003, Peiris/Easther:2006, Peiris/Easther:2006a, Easther/Peiris:2006, Adshead/Easther:2008, Mukherjee/Wang:2005, Leach:2006, Leach/etal:2002, Leach/Liddle:2003, Agarwal/Bean:2009, Powell/Kinney:2007}, interpolating between a finite number of bands \cite{Verde/Peiris:2008, Sealfon/etal:2005, Bridle/etal:2003, Hannestad/Mortsell:2004, Bridges/etal:2006, Bridges/etal:2007, Spergel/etal:2007}, direct inversion of the $C_{\ell}$ spectrum \cite{Shafieloo/Souradeep:2004, Shafieloo/Souradeep:2008, Kogo/etal:2004, Tocchini-Valentini/etal:2005, Nagata/Yokoyama:2008, Nicholson/Contaldi:2009, Nicholson/etal:2010}, or cutting off at finite order some expansion of the inflaton potential $V(\phi)$ \cite{Malquarti/etal:2004, Easther/Kinney:2003, Mortonson/etal:2011}. Unfortunately, the results obtained depend on these assumptions and, for forecasts, also on the input fiducial parameters.

Here we take the inflaton potential to be roughly smooth over scales of order $\sim M_p$, i.e., 
\be
M_p^n \left| \frac{d^n\ln V}{d\phi^n} \right| \lesssim 1\;,  \qquad n=1,2,3 \ldots \;.
\ee
{In the spirit of Refs.~\cite{Stewart:2002, Dodelson/Stewart:2002} we  assume that the first two slow-roll parameters are small. In particular $\epsilon_V = {\cal O}(\lambda)$ and $\eta_V = {\cal O}(\lambda)$ where $\lambda$ is a small parameter which observations -- i.e.~$n_s - 1 \lesssim 0.1$ and $r/16 \lesssim 0.1$ --} suggest is of order $0.1$ or smaller. 
Note that we have not assumed that the running or higher-order slow-roll parameters are small. However, their smallness will be a consequence of our analysis. 

Before proceeding, let us first estimate the excursion of $\phi$, which we call $\delta \phi$, during the $ 9$ $e$-folds corresponding to the observable window  $10^{-4}\,{\rm Mpc}^{-1}\lesssim aH\lesssim \,{\rm Mpc}^{-1}$.  Using the slow-roll conditions, $\delta \phi$ is related to the variation in the number of $e$-folds $\delta N$ by \cite{Lyth:1997}
\be
\delta \phi = M_p \left( \frac{r}{6.9}\right)^{1/2} \delta N\;.
\ee
Using $\delta N = 9$ we obtain $\delta\phi =  3.4  \sqrt{r} M_p$, which implies that for reasonable values of $r$, the potential remains smooth inside all the observable window.

We parameterize the observable piece of the potential by Taylor-expanding it in power series around the value of $\phi=\phipiv$ for which 
$aH = \kpiv \equiv 0.05\,{\rm Mpc}^{-1}$, and for simplicity we take $\phipiv=0$. Thus, 
\begin{equation}
V(\phi)=\sum_{n=0}^{\infty}\left. \frac{V_n}{n!} \right|_{\phi=0} \left(\frac{\phi}{M_p}\right)^n\; , \qquad V_n \equiv M_p^n  \frac{d^nV}{d\phi^n}  \label{eq:piece} \;.
\end{equation}
Without loss of generality, we can take $V_1 <0$ and hence $d\phi/dt>0$. With these definitions our smoothness prior translates into $\amp{V_n/V}\lesssim O(1)$ for each $n$.

Naively,  one could truncate the expansion above at some arbitrary order $n_{\max}$ and use $V_0$, $V_1$, \ldots, $V_{n_{\max}}$ as  parameters to constrain. However, it is important to understand how these parameters enter into the expression of the power spectrum, the spectral index and running, which are the quantities that are to be observed. Using the smallness of $\epsilon_V $ and $\eta_V $, the inflationary power spectrum can be written as \cite{Stewart:2002, Dodelson/Stewart:2002}
\be
\begin{split}
\primsca = &\; \frac{V^3}{12 \pi^2 M_p^4 V_1^2} \Bigg\{ 1 + \left( 3 q_1 - \frac76 \right) \left( \frac{V_1}{V}\right)^2  \\
\\&+ \sum_{p=1}^\infty q_p  \frac{V_{p+1}}{V} \left( \frac{V_1}{V}\right)^{p-1} + {\cal O} ({\lambda}^2) \Bigg\} \;, \label{eq:pot_exp}
\end{split}
\ee
where the potential $V$ and its derivatives $V_n$ are computed at Hubble crossing. Here $q_p$ are known numerical coefficients of order unity \footnote{The coefficients $q_p$ can be given in terms of a generating function by 
$\Sigma_{p=0}^{\infty} q_p x^p = 2^{-x} \cos \left( \frac{\pi x}2 \right) \frac{2 \Gamma (2+ x)}{(1-x) (3-x)}\;$  
}.
Moreover, the spectral index is given by 
\be 
n_s = 1 - 3 \left( \frac{V_1}{V} \right)^2 + 2 \sum_{p=0}^\infty q_p \left( \frac{V_1}{V} \right)^p \frac{V_{p+2}}{V} + {\cal O} ({\lambda}^2)\;,
\ee
while the running is
\be
\alpha_s = -2  \sum_{p=0}^\infty q_p \left( \frac{V_1}{V} \right)^{p+1} \frac{V_{p+3}}{V}+ {\cal O} ({\lambda}^2)\;. \label{eq:alpha_pot}
\ee
(Note that setting $q_p=0$ for $p>0$ we recover the standard slow-roll results of section~\ref{sec:theory}.)

Given the assumption of smoothness introduced above and given that $(V_1/V)^2 = {\cal O}({\lambda})$, terms of order $\sim \frac{V_5}{V} (\frac{V_1}{V})^{3}$ or higher are negligible in eqs.~\eqref{eq:pot_exp}--\eqref{eq:alpha_pot} and can be discarded. In other words, because in slow-roll inflation the excursion of $\phi/M_p$ during 10 $e$-folds is small, a local Taylor expansion truncated at 4th-order will be accurate enough to reproduce the correct power spectrum $\primsca$. The higher-order coefficients in the Taylor expansion are not measurable, unless they are much larger than unity, and they will not be included in our parametrization. 

Thus, in order to derive constraints on the potential we chose parameters that are closely related to those of the familiar CosmoMC, i.e.~$\ln A_s$, $r$, $n_s$ and $\alpha_s$. 
These are
\begin{equation}
\begin{split}
\ln \tilde{A_s} \equiv   \ln \frac{V^3}{12\pi^2M_p^4V_1^2}  \;, \label{eq:tildeAsdef}\qquad
\tilde{r} \equiv   8\left( \frac{V_1}{V} \right)^2 \;, \\
\tilde{n}_s \equiv   1 + 2 \frac{V_2}{V} - 3 \left(\frac{V_1}{V}\right)^2 + 2 q_1 \frac{V_1 V_3}{V^2} + 2q_2 \frac{V_1^2V_4}{V^3}  \;,
\end{split}
\end{equation}
where the right-hand sides of these expressions are evaluated at $\phi=0$,
and
\be
\tilde \alpha_s \equiv \left. -2   \frac{V_1}{V} \frac{V_{3}}{V} -2 q_1 \left( \frac{V_1}{V} \right)^{2} \frac{V_{4}}{V} \right|_{\phi=0} \;, \label{eq:alphas_pot}
\ee
with $q_1 = 1.063$ and $q_2 = 0.209$ \cite{Dodelson/Stewart:2002}. A tilde is used to distinguish the parameters in eqs.~\eqref{eq:tildeAsdef} and \eqref{eq:alphas_pot}, defined in terms of the potential, from the observed quantities $A_s$, $n_s$ and $\alpha_s$ of the power spectrum in eq.~\eqref{eq:shape} and the tensor-to-scalar ratio $r$ in eq.~\eqref{eq:r_ts}.
Finally, we still miss one independent parameter which we take to be proportional to $V_4/V$ and normalize in such a way that it is approximately equal to $V_4/V$ when $r$ is about $0.1$, i.e.
\be
 c_4 \equiv  \left. \frac{V_4}{V} \frac{\tilde r}{0.1} \right|_{\phi=0} \; . \label{eq:c4def}
\ee

\begin{table*}
\caption{Constraints on the inflationary parameters \label{tbl:c3c4}}
\centering
\begin{tabular}{lllll}
\hline
\hline
 & \ fiducial $r=0$ \ &\  fiducial $r=0$ \ & \ fiducial $r=0.128$ \ &\  fiducial $r=0.128$\  \\
 &  Planck  & Planck + LSS pess. &Planck& Planck + LSS pess.  \\
\hline
$\ln(10^{10}\tilde{A}_s)$ & $3.023^{+0.009}_{-0.009}$ \ & $3.022^{+0.008}_{-0.008}$  & $3.010^{+0.010}_{-0.010}$ & $3.007^{+0.009}_{-0.008}$  \\
$\tilde{r}$ & $0.000^{+0.018+0.036}$  & $0.000^{+0.018+0.036}$  & $0.125^{+0.019}_{-0.019}$ & $0.126^{+0.019}_{-0.018}$  \\
$\tilde{n}_s$ & $0.970^{+0.004}_{-0.004}$  & $0.970^{+0.003}_{-0.003}$ & $0.966^{+0.004}_{-0.004}$ & $0.967^{+0.003}_{-0.003}$ \\
$\tilde \alpha_s$ &  $0.000^{+0.005}_{-0.003}$ & $0.000^{+0.004}_{-0.004}$  & $0.002^{+0.005}_{-0.005}$ & $0.013^{+0.004}_{-0.003}$  \\
$c_4$   & $0.02^{+0.22}_{-0.23}$ & $0.01^{+0.22}_{-0.22}$  & $0.15^{+0.34}_{-0.22}$ & $0.11^{+0.25}_{-0.20}$ \\
\hline
\end{tabular}
\end{table*}

\begin{figure}
\centering
\includegraphics[width=0.45\textwidth]{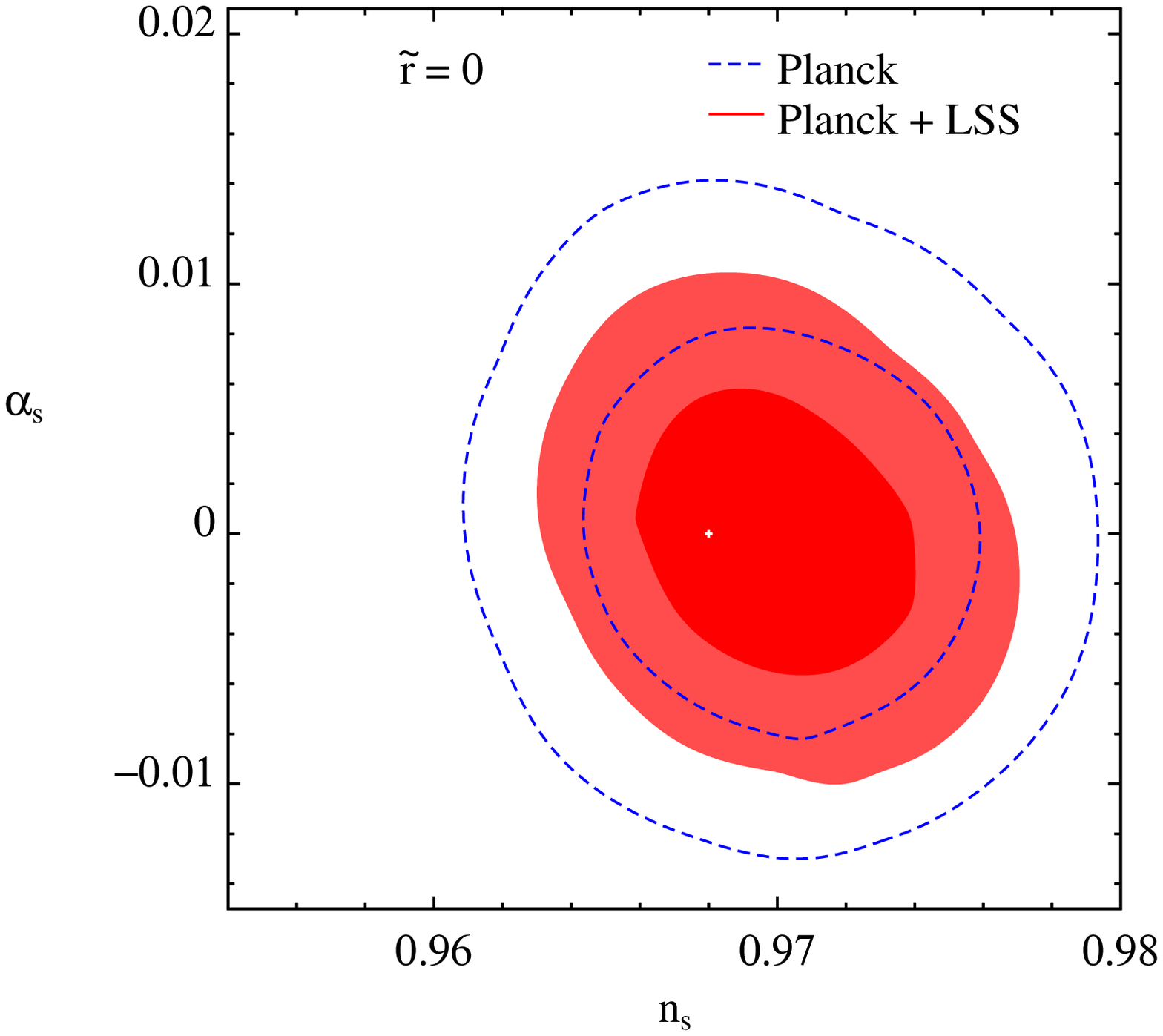}
\includegraphics[width=0.45\textwidth]{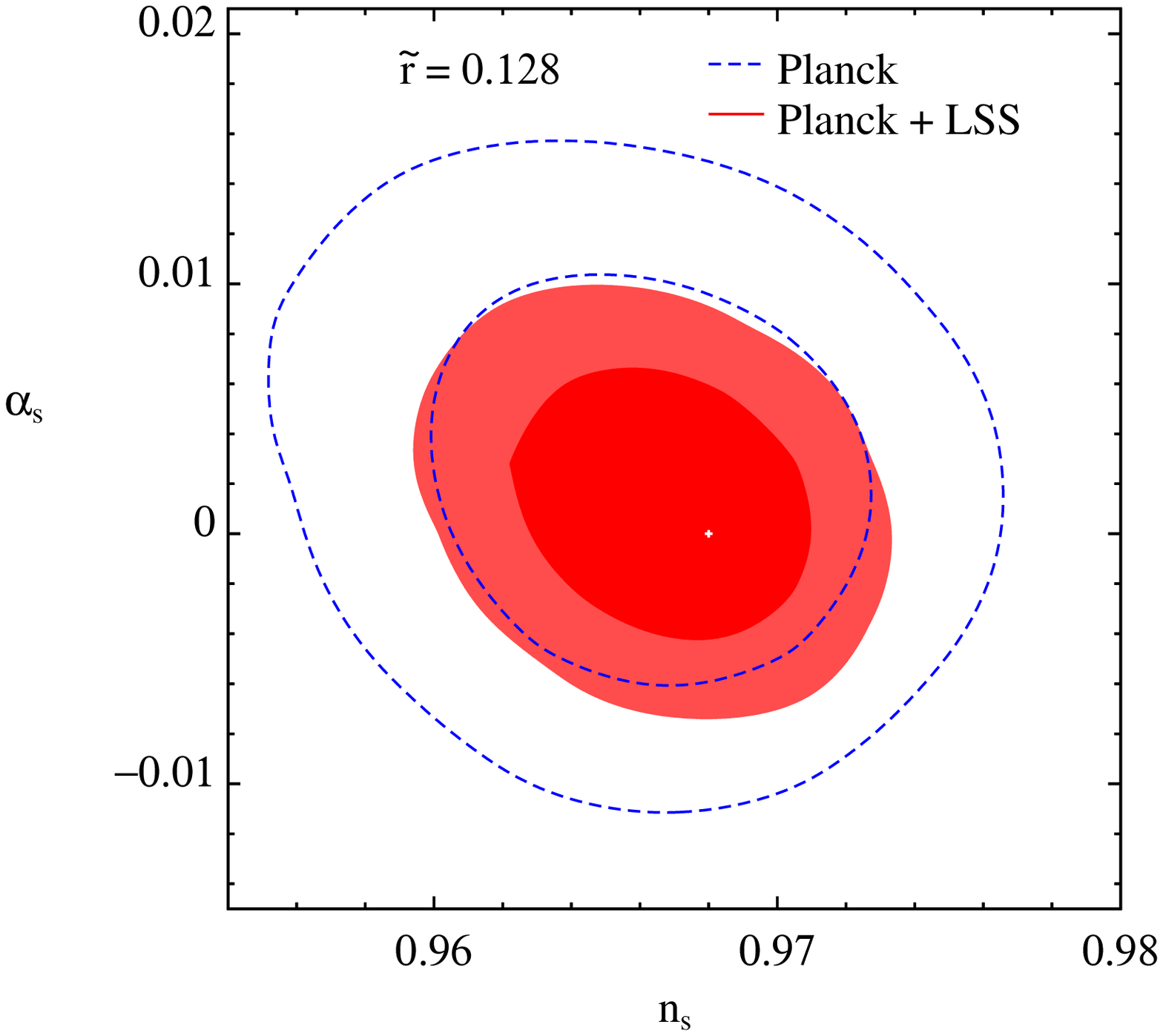}
\caption{68.3\% and 95\% CL contours on the plane $\tilde n_s$-$\tilde \alpha_s$ for Planck alone (blue dashed lines) and for Planck + LSS pessimistic forecast (filled red regions). On the top panel a fiducial $\tilde r=0$ is used, while in the bottom panel the fiducial value of $\tilde r$ is $0.128$ (such as for quadratic inflation). \label{fig:c3c4}}
\end{figure}

We calculate $\primsca$ and $\primten$ for each set of parameters in eqs.~\eqref{eq:tildeAsdef}--\eqref{eq:c4def} by numerically  solving for the evolution of scalar and tensor perturbations $\zeta_k$ and $h_k$ for $10^3$ wavenumbers log-uniformly distributed in $-9 \le \ln [k/{\rm Mpc}^{-1}] \le 1$ and by interpolating in between \footnote{When the slow-roll approximation cannot be applied, we numerically solve for the mode functions $\zeta_k$ and $h_k$.
During inflation, standard Bunch-Davies vacuum initial conditions are chosen for $k \gg aH$. Then, the final spectra are evaluated when all the relevant modes are well outside the horizon, for $k \ll a H$. A better accuracy than $0.1\%$ can be achieved if we  choose values of $aH$ such that ${k}/{aH} > 10^3$ for the initial conditions and ${k}/{aH} < 10^{-3}$ at the end. In our calculations we largely satisfy these criteria.}. Those parameters producing a non-monotonic potential or a non-inflationary phase ($\ddot a<0$) are discarded. 
In particular, we study two examples with fiducial tensor-to-scalar ratio $\tilde r=0$ (top panel) and fiducial $\tilde r=0.128$ (bottom panel of Fig.~\ref{fig:c3c4}).
The 68.3\% CL constraints on the parameters in eqs.~\eqref{eq:tildeAsdef}--\eqref{eq:c4def} are reported in  Table~\ref{tbl:c3c4}. (In the fiducial $\tilde r=0$ case we report the 68.3\% CL and 95.4\% CL upper bounds on $\tilde{r}$.) 
The uncertainties on these parameters -- and on those not reported in this table --  depend little on the fiducial value of $r$ and are slightly worse than those obtained in section~\ref{sec:results_ns} in the case of slow-roll inflation (see Table~\ref{tbl:cosmoparams}), which is expected because we have introduced a new parameter, $c_4$. Thus, the improvement upon the case of Planck alone is again mild. Moreover, the new parameter $c_4$ will not be well constrained by future data. 
In conclusion, with Planck and LSS surveys we will be able to measure $V_3/V$ with an accuracy of $\sim 0.1 \left( \frac{0.1}{r} \right)^{1/2}$ and $V_4/V$ with an accuracy of $\sim 0.5 \left(\frac{0.1}{r}\right)$. Finally, for comparison with the slow-roll case, in Fig.~\ref{fig:c3c4} we show the constraints on the plane $\tilde n_s$-$\tilde \alpha_s$.

A final remark is in order here. Apart from the assumptions of smoothness discussed at the beginning of this subsection, here we also assumed uniform priors on $\ln \tilde{A}_s$, $\tilde{r}$, $\tilde{n}_s$, $\tilde \alpha_s$ and $c_4$.
For a well constrained parameter the choice of prior does not significantly change the results. This is the case, for instance,  for the parameters $\tilde{r}$, $\ln \tilde{A}_s$, $\tilde{n}_s$ and $\tilde \alpha_s$ used in our parametrization, which are closely related to observables. Note, however, that in other approaches to potential reconstruction, such as the flow-equation approach \cite{Kinney:2002}, the cutoffs of the expansion and the priors/boundaries of parameters are often chosen arbitrarily.

\section{Constraining features in the potential \label{sec:glitches}}

In this section we study inflationary models that predict features in the primordial scalar power spectrum. The amplitude of these features is small compared to the overall amplitude of the spectrum, typically of a few percent or less. 
Before dealing with concrete models, let us estimate the typical width in momentum space that can be measured by a perfect large-scale structure  and CMB experiment. A similar discussion can be found in \cite{Hamann/etal:2008}.
We will confirm these expectations more precisely in the following subsection when studying concrete models.

\subsection{Analytical estimates} \label{sec:analytical}

\subsubsection{Large scale structure survey}
\label{sec:LSS_an}

For large-scale structure, the main limitation comes from the finite size of the survey. Indeed, the fact that the survey volume is finite sets a coherence wavelength $k_{\rm min} \sim 2 \pi/V^{1/3}$, where $V$ is the volume of the survey, below which all features get smeared \cite{Baumgart/Fry:1991}.  Practically, the density field gets smoothed by a window function $W$ determined by the survey geometry; in Fourier space this smoothing window is typically a compact ball with width of order $k_{\rm min}$. For simplicity, let us  neglect {in this discussion} the effect of shot noise and assume that the smoothing window  is simply a Gaussian function with half width $\sigma_W$. In this case, for each wavemode $k$ the estimated power spectrum is given by
\be
\hat{ P}(k) = \int \frac{d^3 k'}{(2 \pi)^3} P(k') | W ( | \vk - \vk ' |)|^2\; , \label{estimator_BF}
\ee
where
\be
|W(k)|^2 = \frac{1}{\sqrt{2 \pi} \sigma_W  } \exp \left(- \frac{k^2}{2 \sigma^2_W} \right) \;.  \label{window_gauss}
\ee
We take the half width of the Gaussian to be exactly equal to the  coherence wavelength $k_{\rm min}$, $\sigma_W = k_{\rm min}$.

Let us study the effect of the window function on the estimator in the case of a power spectrum with features. As a warm up, we first consider a Gaussian bump of {relative} amplitude $\A$, centered at $k=k_B$ and of half width $\sigma_B$, i.e.
\be
P(k) = P_0(k) \left[1 + \A \cdot \exp \left(- \frac{(k-k_B)^2}{2 \sigma^2_B} \right)  \right] \;, \label{PS_bump}
\ee
where $P_0(k)$ is a smooth power spectrum.
The estimated power spectrum can be computed by plugging this expression into eq.~\eqref{estimator_BF}. In the limit where the smooth power spectrum varies little in an interval $\sigma_W$ or $\sigma_B$, one can easily compute the estimated power spectrum analytically
{and the calculation can be simplified by considering} wavemodes $k \gg \sigma_B$.
{As expected, for a window function much smaller than that of the bump, $\sigma_W \ll \sigma_B$, the estimator in eq.~\eqref{PS_bump} reproduces the original power spectrum \eqref{PS_bump}.}
In this case the condition of observability of the bump is simply given by 
$\A \gtrsim N_B^{-1/2}$, where $N_B \equiv 4 \pi k_B^2 \sigma_B V /(2 \pi)^3 =  4 \pi k_B^2 \sigma_B /k_{\rm min}^3 $ is the number of modes inside  a shell of radius $k_B$ and width $\sigma_B$. By defining the relative  width of the bump $\delta \ln k_B \equiv \sigma_B / k_B$, this relation can be rewritten as
\be
\A \cdot \sqrt{\delta \ln k_B} \gtrsim  \left( \frac{k_{\rm min}}{k_B} \right)^{3/2} \gtrsim  10^{-3}\;, \label{feat_exp}
\ee
where in the last inequality we have used the typical coherence frequency  of galaxy surveys such as the one considered here, i.e.~$k_{\rm min} = 10^{-3}\, h/ \text{Mpc}$ and $k_B \lesssim k_{\rm max} = 0.1 \, h /\text{Mpc} $.
In the opposite case, i.e.~when the bump is much smaller than the coherence length, $\sigma_B \ll \sigma_W$, one obtains
\be
\hat P(k) \simeq P_0(k) \left[1 + \A \cdot \frac{\sigma_B }{\sigma_W } \exp \left(- \frac{(k-k_B)^2}{2 \sigma^2_W} \right)  \right] \;. \label{PS_bump2}
\ee
Thus,  the amplitude of the bump in the estimated power spectrum is suppressed by a factor $\sigma_B/\sigma_W$ while its width gets 
smeared out to  $\sigma_W$. Although the bump cannot be resolved, it can 
be measured  if $\A  \gtrsim (\sigma_W/\sigma_B)^{1/2}  N_B^{-1/2} \gg N_B^{-1/2}$. Using $\sigma_B/k_B = \delta \ln k_B$, we can rewrite this relation as
\be
\A \cdot \delta \ln k_B \gtrsim   \left( \frac{k_{\rm min}}{k_B} \right)^2 \gtrsim  10^{-4}\;.
\ee

Consider now the case of a power spectrum with periodic oscillations of frequency $\sigma_B$, i.e.
\be
P(k) = P_0(k) \left[1 + \A \cdot \cos ({k/ \sigma_B} )  \right] \;. \label{PS_oscill}
\ee
Using eqs.~\eqref{estimator_BF} and \eqref{window_gauss} we can compute the estimated power spectrum similarly to what done above. For frequency $\sigma_B \gg \sigma_W$ one recovers eq.~\eqref{PS_oscill}, while in the opposite case, $\sigma_B \ll \sigma_W$, one obtains
\be
\hat P(k) = P_0(k) \left[1 + \A \cdot \exp\left({- \frac{\sigma_W^2}{2 \sigma_B^2}} \right) \cos ({k/ \sigma_B} )  \right] \;. \label{PS_oscill2}
\ee
The frequency and phase of oscillations do not change but {their} amplitude  gets exponentially suppressed for $\sigma_B \ll \sigma_W$. The condition for observability is now obtained by comparing the estimated amplitude of the oscillations with $N^{-1/2}_{\rm tot}$, where $N_{\rm tot}$ is the total number of modes in the survey, i.e.~$N_{\rm tot} \sim ({k_{\rm max}}/{k_{\rm min}})^3$. This then becomes 
\be
\A \cdot \exp\left({- \frac{\sigma_W^2}{2 \sigma_B^2}} \right) \gtrsim   \left(\frac{k_{\rm min}}{k_{\rm max}} \right)^{3/2} \sim  10^{-3}\;. \label{LSS_an}
\ee

In practice, to implement in our analysis the fact that there is a fundamental coherence wavemode below which we cannot resolve, we have applied the following prescription. For each redshift bin we have sampled the wavenumber $k$ in units of $k_{\rm min}$ ($k_{\min} ( z) =  2\pi/V^{1/3}( z)$). For each (discrete) value of $k$ we have computed the power spectrum using eqs.~\eqref{estimator_BF} and \eqref{window_gauss}, where the width of the Gaussian window $\sigma_W$ has been chosen to be 
\be
\sigma_W = \frac{\sqrt{2 \ln 2}}{2 \pi} \left(\frac{4 \pi}{3} \right)^{1/3} k_{\rm min} \simeq 0.302 \; k_{\rm min}\;. 
\ee
In such a way, the real-space representation of the window function, if cut off at its half-height, contains the same volume
as that of the redshift bin. 
The fact that  $\sigma_W$ is smaller than $k_{\rm min}$ allows us to neglect the overlap between window functions centered around   neighboring values of $k$ but slightly  weakens the constraints discussed above.

To summarize, using LSS data we will be able to resolve features larger than roughly the coherence wavemode of the survey, $k_{\rm min}$, corresponding to a relative width  $\delta \ln k \gtrsim 10^{-2}$, whose amplitude is at least of order $(k_{\rm min}/k_{\rm max})^{3/2} \sim 10^{-3}$.

\subsubsection{Cosmic Microwave Background}
\label{sec:CMB}
{The coherence wavemode of the CMB, i.e.~the smallest $k$ mode probed by the temperature anisotropies, roughly corresponds to $k_{\min} \sim D_{\rm rec}^{-1}$, where $D_{\rm rec}$ is the comoving distance to the last-scattering surface. More specifically, $D_{\rm rec} \equiv \tau_0 - \tau_{\rm rec}$, where $\tau_0$ and $\tau_{\rm rec}$ are the conformal time today and at recombination, respectively. Thus, we expect that features in the primordial power spectrum with a width in momentum space smaller than  $D_{\rm rec}^{-1}$ cannot be measured.}
This implies that we can only resolve features with relative width such that $\delta \ln k \gtrsim k_{\rm min}/k_{\rm max} \sim \ell^{-1}_{\rm max} $, where $\ell_{\rm max}$ is the largest multipole of the CMB map. However, this naive estimate is too optimistic. Indeed, the projection from momentum space to multipole space degrades the amplitude of features, hampering their measurement. The order of magnitude of this effect can be schematically computed by the following example.

To simplify things as much as possible, let us consider the angular spectrum of the temperature anisotropies in the Sachs-Wolfe, instantaneous recombination and flat-sky limit. This is given by 
\be
C_\ell = \frac{1}{18 \pi D^2_{\rm rec}} \int_{-\infty}^{+\infty} d y \; P(k=\sqrt{\ell^2/D_{\rm rec}^2 + y^2}) \;.
\label{CL_oscil}
\ee
Plugging the expression for the power spectrum with oscillations, eq.~\eqref{PS_oscill}, into this equation, one can solve the integral on the right-hand side for a scale invariant spectral index of primordial fluctuations. Its solution can be written in terms of a Meijer G-function \footnote{The exact solution in terms of a Meijer G-function is 
\be
C_\ell = \frac{\Delta^2}{9 \pi \ell^2}  \left[1+ \delta \ln A_s \frac{\pi}{2} G_{1\, 3}^{2\, 0} \begin{pmatrix}
\frac{x^2}{4} \; \begin{array}{|c }
  3/2 \\
  0 , 1 , 1/2
 \end{array}
  \end{pmatrix} \right]\;,
\ee
where $x \equiv \ell/(\sigma_B D_{\rm rec})$ and we have used $P(k) = \Delta^2 k^{-3}$.
}, {but it is more enlightening to use its asymptotic expansion in the large $x \equiv \ell/\sigma_B D_{\rm rec}$ limit, by which we can recast eq.~\eqref{CL_oscil} as \cite{Adshead/etal:2011b}}
\be
C_\ell \simeq C_{0,\ell} \left[ 1 + \A  \left( \frac{\pi }{ 2 x (\ell)} \right)^{1/2} \cos \left( x(\ell) + \frac{\pi}{4} \right) \right]\;,
\ee
where $C_{0,\ell}$ is the angular CMB spectrum for a smooth initial power spectrum. Thus, apart from a phase shift, the oscillatory behavior does not change. However, the amplitude of the oscillations gets suppressed by a factor $\sim (\sigma_B D_{\rm rec} /\ell)^{1/2}$ \cite{Adshead/etal:2011b}. 
For  oscillations with constant frequency in $k$ the suppression factor is proportional to $\ell^{-1/2}$ and hence becomes more important at large $\ell$. Note however that for  oscillations with constant frequency in $\ln k$, such as in axion monodromy, the suppression factor is independent of $\ell$ and, using $\sigma_B  \simeq  (\ell / D_{\rm rec}) \delta \ln k  $, it is simply given by  $\sqrt{\delta \ln k}$.

{To measure these oscillations,  their relative amplitude in the $C_\ell$ must be larger} than the inverse square-root of the number of pixels. From the discussion above this is given by
\be
\A \cdot \sqrt{\delta \ln k} \gtrsim 5 \cdot 10^{-4}\;, \label{CMB_const}
\ee
where we have used  $l_{\rm max} = 1500$ (for larger $\ell$'s, the noise is larger than the signal). 

As shown in Ref.~\cite{Adshead/etal:2011b}, there is also a smearing effect due to gravitational lensing starting at $\ell \sim 10^3$. As most of our constraints come from $\ell \lesssim 10^3$, where CMB lensing is negligible, we did not include this effect in our analysis.

For the analysis that will be discussed in {the rest of} this section, the presence of small features and oscillations makes the calculation of the angular power spectrum 
challenging. Indeed, in order to speed up the computation,  in  publicly available CMB codes such as CAMB \cite{Lewis/etal:2000},  CLASS \cite{Lesgourgues:2011, Blas/etal:2011} or CMBEasy \cite{Doran:2005}, the $C_\ell$'s are computed by sampling the primordial power spectrum with steps longer than the width of these features and using a certain number of approximations which in our case are not accurate enough. For this reason, one of the authors {of this article} (ZH) has developed a new CMB Boltzmann code which is adapted to treat very sharp features in the primordial  power spectrum at {a} reasonable computing time. This code is written in {the} Newtonian gauge and agrees well with other codes such as CAMB and CLASS.  The code is presented in \cite{Huang:2012} but we  outline {its main features in the appendix, Sec.~\ref{sec:tech}.}

\subsection{Ringing features} \label{sec:ringing}

{As an example of model of inflation predicting a sharp feature in the power spectrum we consider the model first presented by Starobinsky in \cite{Starobinsky:1992}. The feature is produced by a discontinuity in the first derivative of a linear potential. More precisely, this} can be defined by its amplitude at $\phi=\phipiv$ (the value of the field at Hubble-exit of the pivot scale $\kpiv$),  i.e.~$V_* \equiv V(\phipiv ) $, and its first derivative given by
\begin{equation}
 M_p\frac{dV}{d\phi} =\left\{
\begin{array}{ll} 
V_{1, - } \quad &\text{ if } \phi <\phi_\pm\ , \\
V_{ 1,+} \quad &\text{ if } \phi >\phi_\pm\ , 
\end{array}
\right. \label{eq:step}
\end{equation}
where $V_{1,-}$ and $V_{1,+}$ are constants. A closed form of the potential can be obtained by integrating $dV/d\phi$ from $\phi=\phipiv$. With this definition the inflaton mass $d^2V/d\phi^2$ diverges at $\phi = \phi_\pm$, which is theoretically fine but numerically problematic.  Thus, in practice we take a slightly modified potential which smoothly interpolates the jump of the first derivative: we use eq.~(\ref{eq:step}) for $\amp{\phi-\phi_\pm}\frac{2V_*}{V_{1,-}+V_{ 1,+}}\ge \varepsilon M_p$ and we linearly interpolate $dV/d\phi$ for $\amp{\phi-\phi_\pm}\frac{2V_*}{V_{1,-}+V_{ 1,+}}<\varepsilon M_p$, where $\varepsilon$ is a small parameter. In the example below we consider $\varepsilon=0.01$. We have verified that  $\primsca$ and $\primten$ are stable to changes of $\varepsilon$ around this value and that they do not depend on the choice of $\varepsilon$ as long as $\varepsilon \ll 1$. 

We constrain the following parameters, constructed from $V_*$, $V_{1,-}$, $V_{1,+}$ and $\phi_\pm$,
\begin{align}
\ln \tilde{A_s}  \equiv & \; \ln \frac{V_*^3}{6\pi^2M_p^4(V^2_{ 1,-}+V^2_{1,+})} \;, \label{eq:stepAsdef1}  \\
\tilde{n}_s  \equiv & \; 1 - \frac{3}{2}\Bigg[ \left(\frac{V_{1,-}}{V_*}\right)^2 + \left(\frac{V_{1,+}}{V_*}\right)^2 \Bigg]\;, \label{eq:stepAsdef2}
\end{align}
which approximate the average of $A_s$ and $n_s$, respectively, and
\begin{align}
\delta {n}_s \equiv & \; \frac{3}{2} \left[ \left(\frac{V_{1,-}}{V_*}\right)^2 -\left(\frac{V_{1,+}}{V_*}\right)^2 \right]\;, \label{eq:stepdeltalnk1} \\
\lnkring  \equiv  & \; \phi_\pm \sqrt{\frac{3}{1-\tilde{n}_s}} \; , \label{eq:stepdeltalnk2}
\end{align}
which parameterize the variation of the spectral index and the position where the ringing feature starts. As above, the tilde is used to distinguish  quantities defined by the potential from those defined directly in terms of observables.
For each set of {parameters given in} eqs.~\eqref{eq:stepAsdef1}--\eqref{eq:stepdeltalnk2} we solve $V_*$, $V_{1,-}$, $V_{1,+}$ and $\phi_\pm$ and calculate scalar and tensor perturbations $\zeta_k$ and $h_k$ for $10^3$ wavenumbers log-uniformly {distributed} in $-9\le \ln [k/{\rm Mpc}^{-1}] \le 1$. 

\begin{figure}
\centering
\includegraphics[width=0.5\textwidth]{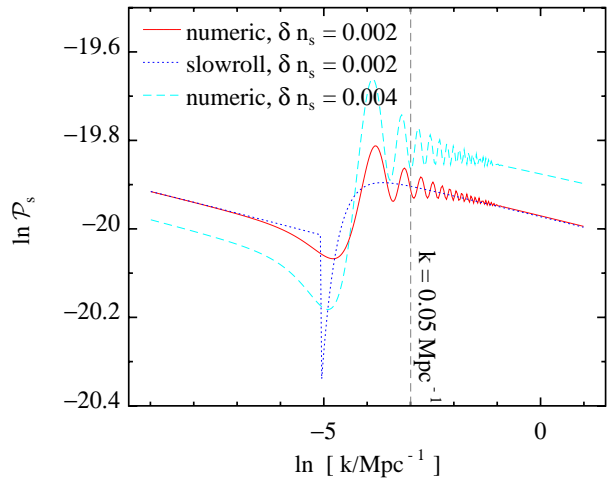}
\caption{The primordial power spectrum $\ln \primsca$, computed by numerically solving the scalar perturbation equations, for a step in $dV/d\phi$ in the Starobinsky model.  
The solid red line is the power spectrum for the model with $\ln \tilde{A_s}=3.02$, $\tilde{n}_s=0.975$, $\delta {n}_s = 0.002$ and $\lnkring = -2$; the dotted line is the slow-roll approximation for the same model. The dashed cyan line is the power spectrum computed for a different value of $\delta n_s$, i.e.~$\delta {n}_s = 0.004$. \label{fig:steplnps}}
\end{figure}

As an example, in Fig.~\ref{fig:steplnps} we show the  scalar power spectrum calculated by numerically solving the evolution equations for scalar perturbations. The solid red line corresponds to  the case $\ln \tilde{A_s}=3.02$, $\tilde{n}_s=0.975$, $\delta {n}_s = 0.002$ and $\lnkring = -2$, which will be studied below. The slow-roll approximation \cite{Stewart/Lyth:1993} is also shown for comparison and as a check of the numerical accuracy. To see the effect of varying $\delta{n}_s$ on the power spectrum,  in the same plot we also show the case for $\delta{n}_s = 0.004$ in cyan.  Typically, in this model the relative width of the feature is of order 1. Its exact value is related to $\delta {n}_s$, but as long as we consider only small values of $\delta {n}_s$ it is not too sensitive to it. Moreover, using eqs.~\eqref{eq:stepAsdef1}--\eqref{eq:stepdeltalnk1}, we can relate the relative jump in the amplitude of the power spectrum to $\delta n_s$, i.e.
\be
\A = \frac{\delta A_s}{\tilde A_s} = \frac{2\delta n_s}{1 - \tilde n_s} \;. \label{ampli_ns}
\ee 
In the case of the example above one finds $\A= 0.16$, which is confirmed by the numerical calculation.

\subsubsection{Null test}

Let us first find the  upper bound on $\delta {n}_s$ that can be obtained in absence of a ringing feature, i.e.~by choosing the fiducial $\delta n_s =0$. For this null test we produce mock data using a fiducial linear potential model with $\ln(10^{10} \tilde{A}_s)=3.02$ and $\tilde{n}_s = 0.975$. 
As a prior on $\lnkring$ we take $-4 \le \lnkring \le 2$, which roughly corresponds to the range of scales probed by Planck and the LSS galaxy survey, i.e.~$10^{-3}\;{\rm Mpc}^{-1} \le k \le0.37\; {\rm Mpc}^{-1}$. Note that in the null-test case one has to chose this prior with care. Indeed, with a bad choice of priors the posterior probability may be dominated by a fraction of the volume of the parameter space which is beyond the observational window and thus insensitive to the data.

\begin{figure}
\includegraphics[width=0.45\textwidth]{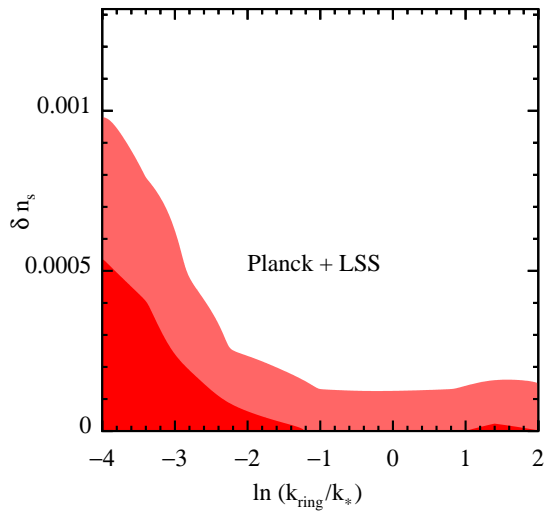}
\caption{The marginalized 68.3\% CL and 95.4\% CL constraints on $\delta{n}_s$ {as a function of $\lnkring$} for ringing features in the Starobinsky model. The mock data are produced using a fiducial linear potential model with $\ln(10^{10} \tilde{A}_s)$, $\tilde{n}_s = 0.975$, and no feature ($\delta {n}_s= 0$). The prior on $\lnkring$ is $-4 \le \lnkring \le 2$. \label{fig:stepnull}}
\end{figure}

We show the constraints using CMB + LSS in Table~\ref{tbl:stepnull} and the $\lnkring$-$\delta n_s$ plane in Fig.~\ref{fig:stepnull}. The standard cosmological parameters such as $\Omega_bh^2$, $\Omega_ch^2$, $\theta$ and $\tau_{\rm re}$ are not affected by the ringing feature, so that choosing a model with an incorrect value of $\delta{n}_s$ or $\lnkring$
has almost no impact on these parameters.  The upper bound of $\delta{n}_s$ depends on $\lnkring$. Indeed, since on smaller scales more Fourier modes are probed by CMB and large-scale structure, the constraint on $\delta {n}_s$ is better for larger $\lnkring$, as shown in Fig.~\ref{fig:stepnull}. It is necessary to use both CMB and  large-scale structure to cover the wide range of scales that we have chosen in the prior. Using only one of them would lead to posteriors dominated by an unconstrained region in the parameter space. Thus, in this case we do not compute the CMB-only constraints.

\begin{table}
\centering
\caption{Null-test  on Starobinsky model \label{tbl:stepnull}}
\begin{tabular}{ll}
\hline
\hline
Parameter & Planck + LSS pess. \\
\hline
$\Omega_bh^2$ & $0.0220^{+0.0001}_{-0.0001}$ \\
$\Omega_ch^2$ & $0.11280^{+0.00024}_{-0.00023}$ \\
$\theta$ & $1.0462^{+0.0002}_{-0.0002}$ \\
$\tau_{\rm re}$ & $0.090^{+0.003}_{-0.003}$ \\
$\tilde{n}_s$ & $0.975^{+0.002}_{-0.002}$ \\
$\lnkring$ & $\text{unconstrained}$ \\
$\ln(10^{10}\tilde{A}_s)$ & $3.013^{+0.009}_{-0.011}$ \\
$\delta n_s$ & $0.0000^{+0.00025+0.00066}$ \\
\hline
\end{tabular}
\end{table}

\subsubsection{Measuring the ringing feature}

Now we study how well CMB and large-scale structure can detect a ringing feature {and as fiducial values we choose} $\delta {n}_s = 0.002$ and $\lnkring = -2$, corresponding to the case plotted in Fig.~\ref{fig:steplnps}. The fiducial values on the other parameters are chosen as in the null-test case.

\begin{figure}
\centering
\includegraphics[width=0.45\textwidth]{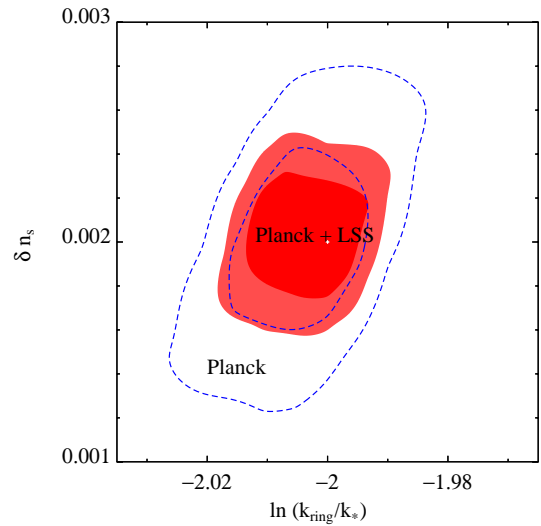}
\caption{The 68.3\% CL and 95.4\% CL constraints on $\delta {n}_s$ and $\lnkring $ with Planck-only (blue) and Planck plus LSS survey pessimistic forecast (red),  for ringing features in the Starobinsky model. The white point is the fiducial model. \label{fig:deltanslnk}}
\end{figure}

The forecasted constraints on the cosmological parameters are listed in Table~\ref{tbl:step}. The constraints on the parameters not listed in this table are very close to those for the null case, listed in Table~\ref{tbl:stepnull}. 
The  constraints on the $\lnkring$-$\delta{n}_s$ plane are shown in Fig.~\ref{fig:deltanslnk}.  As we explained in the previous subsection, when mapping the power spectrum to the angular $C_\ell$, the narrowest features in the power spectrum get suppressed. Thus, the CMB experiment is likely to capture only the first two or three oscillations of the ringing features while a galaxy survey captures more oscillations. Indeed, the LSS data  improve the total FOM for $\delta {n}_s$, $\lnkring$, $\tilde{n}_s$ and $\ln \tilde{A}_s$ by a factor of about $5$. 
\begin{table}
\centering
\caption{Constraints on Starobinsky model \label{tbl:step}}
\begin{tabular}{lll}
\hline
\hline
& Planck only & Planck + LSS pess.  \\
\hline
$\tilde{n}_s$  & $0.975^{+0.003}_{-0.003}$ & $0.975^{+0.002}_{-0.002}$ \\
$\lnkring$ & $-2.000^{+0.010}_{-0.010}$  & $-2.004^{+0.007}_{-0.007}$ \\
$\ln(10^{10}\tilde{A}_s)$  & $3.021^{+0.011}_{-0.010}$ & $3.022^{+0.008}_{-0.008}$\\
$\delta {n}_s$  & $0.00200^{+0.00032}_{-0.00033}$ & $0.00203^{+0.00020}_{-0.00020}$ \\
\hline
\end{tabular}
\end{table}
Note that the constraints for LSS can be qualitatively derived using eq.~\eqref{feat_exp}. Indeed, using that the relative amplitude of the features is of order 1, this equation and eq.~\eqref{ampli_ns} imply that the typical constraint on $\delta n_s$ is of order $\sigma(\delta n_s)\sim (k_{\rm min}/k_{\rm ring})^{3/2} (1-n_s)$. By plugging the numbers in we find $\sigma(\delta n_s) \sim 0.0015$, which confirms the numerical results.

\subsection{Constraining oscillations: axion monodromy} \label{sec:monodromy}

For the axion monodromy model we use the parameterized potential given in Ref.~\cite{Flauger/etal:2010},
\begin{equation}
V(\phi) = \mu^3\left[\phi + bf\cos\left(\frac{\phi}{f}\right)\right]\;, \label{eq:monodromy}
\end{equation}
where $\mu$ parameterizes the scale of inflation, $b$ is a small parameter (monotonicity of the potential requires $b \ll 1$), $f$ is a -- typically sub-Planckian -- frequency, $f\ll M_p$, and inflation requires $\phi \gg M_p$. In this model inflation ends when $\phi \simeq 0$. As usual, $\phipiv$ denotes the value of the field when the pivot scale exits the Hubble radius, which for 60 $e$-folds of inflation is  $\phipiv \simeq 11M_p$. 

Let us now map a variation of $f$ in field space, $\delta\phi = f$, into the corresponding variation in momentum space. We work at lowest order in the slow-roll approximation, where $k \simeq e^{H t} H$ with $H$ constant. In this case one finds
\be
\delta \ln k  =  \frac{ H}{\dot \phi}  \delta \phi  =\frac{ H}{\dot \phi}  f  \;. \label{k-phi}
\ee
Furthermore, neglecting the oscillations of the potential in the Klein-Gordon equation of the field $\phi$ one obtains $3 H \dot \phi = - \mu^3$ and replacing $H$ using the Friedmann equation ($3 M_p^2 H^2 = \mu^3 \phi$) one gets $ {H}/{\dot \phi} = - {\phi}/{M_p^2}$. With $\phipiv$ as typical value for the scalar field $\phi$ we obtain the width of oscillation of the power spectrum in $\ln k$,
\begin{equation}
 \delta {\ln k} = \frac{ f\phipiv}{M_p^2} \; , \label{freq_k}
\end{equation}
which  will be used as a parameter to constrain.

We parameterize the primordial power spectrum using the analytical approximation given in Ref.~\cite{Flauger/etal:2010},
\begin{equation}
   \primsca(k) = A_s \left(\frac{k}{\kpiv}\right)^{n_s-1}\left[1 + \delta n_s \cos \left(\frac{\phi_k}{f}  + \Delta \varphi \right)\right]\; , \label{eq:monoPs2}
\end{equation}
where 
\be
\delta n_s = \frac{12 b}{\sqrt{1+ \left(\frac{3 f \phipiv}{M_p^2}\right)^2}} \sqrt{\frac{\pi}{8} \coth \left( \frac{\pi M_p^2}{2 f \phipiv} \right) \frac{f \phipiv}{M_p^2}}\;. \label{eq:monodeltans}
\ee
The phase $\Delta \varphi$ originates from the uncertainty both in the exact number of $e$-fold from the end of inflation and, on a more microscopical level, in the exact sinusoidal modulation of the scalar potential arising in the string theory construction.

By integrating the first equality of eq.~\eqref{k-phi} {using again $ {H}/{\dot \phi} = - {\phi}/{M_p^2}$} one obtains \cite{Flauger/etal:2010}
\be
\phi_k = \sqrt{\phipiv^2 - 2 \ln k/k_*} \simeq \phipiv - \frac{\ln k/k_*}{\phipiv}\;,
\ee
which can be plugged into the argument of the cosine in eq.~\eqref{eq:monoPs2}. Rewriting this argument using eq.~\eqref{freq_k}, {eq.~\eqref{eq:monoPs2} becomes}
\begin{equation}
   \primsca(k) = A_s  \left(\frac{k}{\kpiv}\right)^{n_s-1}\left[1 + \delta n_s \cos \left(\frac{\ln k/k_*}{\delta\ln k} + \varphi  \right)\right]\; , \label{eq:monoPs}
\end{equation}
where the phase $\varphi$ is given by $\varphi = -\Delta \varphi - \phipiv/f$. 
We have verified that Eq.~(\ref{eq:monoPs}), which does not assume the slow-roll approximation, for $b\ll 1$ and $\delta\ln k\gtrsim 0.003$ agrees to the percent level  with the power spectrum obtained by numerically  solving the scalar perturbation equations during inflation.

In the following we will constrain 5 parameters. Two of them, the width of the oscillations $\delta \ln k$, defined in eq.~\eqref{k-phi}, and the free phase $\varphi$, have been already introduced.  We allow the phase $\varphi$ to vary between $-\pi$ and $\pi$.
We will then constrain the amplitude and spectral index of the oscillations, which are defined in terms of the physical parameters $\phipiv$ and $\mu^3$ as
\be
{\tilde A_s} =   \frac{\mu^3\phipiv^{3}}{12\pi^2M_p^6} \;, \label{eq:monoAs}\qquad
{\tilde n_s} =  1 - \frac{3M_p^2}{\phipiv^{2}} \;,
\ee
and the deviation from the spectral index $n_s$, which is related to the physical parameter $b$ and $f$ in the potential by eq.~\eqref{eq:monodeltans}.
For the MCMC calculation we assume uniform priors on these five parameters. For the tensor spectrum we use that, at lowest-order in the slow-roll approximation, $r = (8/3)(1-n_s)$ and $n_t = -{r}/{8}$.

\subsubsection{Null test}

\begin{figure}
  \includegraphics[width=0.45\textwidth]{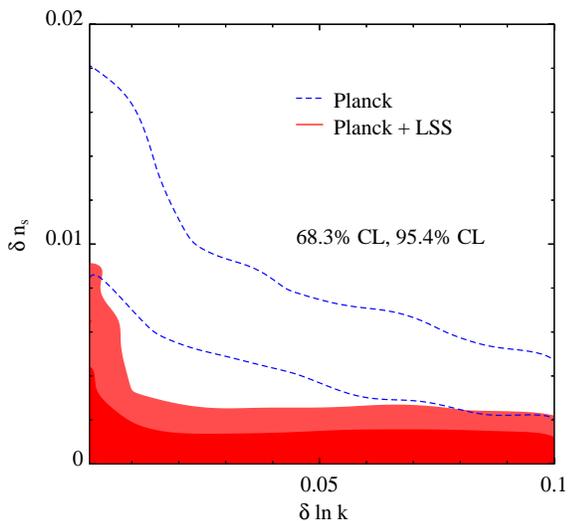}%
  \caption{The marginalized posterior 68.3\%CL and 95.4\% CL constraints on $\delta n_s$ and $\delta\ln k$ for Planck-only (blue) and Planck plus a LSS survey (red). Note that there are some unavoidable resolution effects in this representation of the constraints:  the 1 and 2-$\sigma$ contour lines should go  infinity for $\delta \ln k \to 0$, as for very small values of $\delta \ln k$ one does not have any constraints on oscillations.
The fact that  contour lines remain finite is due to a practical choice of finite range for $\delta \ln k$: {our prior on $\delta \ln k$ is $0.001 \le \delta \ln k \le 0.1$ (for $\delta n_s$ we have taken $0 \le \delta n_s \le 0.05$).} This problem disappears, of course, in the case of a  detection. \label{fig:mononull}}
\end{figure}

\begin{table}
\centering
\caption{Null test on axion monodromy\label{tbl:mononull}}
\begin{tabular}{lll}
\hline
\hline
Parameter & Planck only & Planck + LSS pess. \\
\hline
$\Omega_bh^2$ & $0.02200^{+0.00011}_{-0.00011}$ & $0.021996^{+0.000085}_{-0.000074}$ \\
$\Omega_ch^2$ & $0.1128^{+0.0009}_{-0.0009}$ & $0.11279^{+0.00024}_{-0.00023}$ \\
$\theta$ & $1.04618^{+0.00021}_{-0.00022}$ & $1.04618^{+0.00019}_{-0.00018}$ \\
$\tau$ & $0.090^{+0.004}_{-0.004}$ & $0.090^{+0.003}_{-0.003}$ \\
$\ln(10^{10} \tilde A_s)$ & $3.027^{+0.008}_{-0.008}$ & $3.0268^{+0.0063}_{-0.0057}$ \\
$\tilde n_s$ & $0.9749^{+0.0023}_{-0.0025}$ & $0.9751^{+0.0018}_{-0.0018}$ \\
$\delta  n_s$ & $0.0000^{+0.0046+0.0132}$ & $0.0000^{+0.0015+0.0046}$ \\
\hline
\end{tabular}
\end{table}
As before, we start by considering the null test.
Mock data is generated using a fiducial power-law spectrum with ${\tilde n_s}=0.975$ and $\ln (10^{10} {\tilde A_s}) = 3.027$. The constraints on the parameters are given in Table~\ref{tbl:mononull}. {As the frequency $\delta \ln k$ and phase $\varphi$ remain unconstrained, they are not shown in this table.} The constraints in the plane $\delta \ln k$-$\delta n_s$ are shown in Fig.~\ref{fig:mononull}. As expected, in the Planck-only case the limit on $\delta n_s$ depends on the frequency $\delta \ln k$. For $\delta \ln k =0.1$ we find  $\delta n_s \lesssim 0.2$ at 68.3\% CL, which is about what expected from eq.~\eqref{CMB_const}. This constraint weakens at smaller $\delta \ln k$. Adding a large-scale structure survey  with specifications such as those considered here improves these constraints, which for  $\delta \ln k \gtrsim 0.01$ are independent of $\delta \ln k$, as expected from the discussion of Sec.~\ref{sec:LSS_an}. Indeed, when $\delta \ln k $ becomes smaller than {$k_{\rm min}/k_{\rm max}$, which roughly corresponds to the relative} coherence length of the survey, the amplitude of oscillations of the estimated power spectrum remains unconstrained. 

Note that one should be aware of prior-dependence when interpreting the contours shown in Fig.~\ref{fig:mononull}.  For instance, one would expect to see the 1 and 2-$\sigma$ contour lines to go to  infinity for $\delta \ln k \to 0$, as for very small values of $\delta \ln k$ one does not have any constraints on oscillations.
The fact that  contour lines remain finite is due to a practical choice of finite range for $\delta \ln k$, i.e.~$0.001 \le \delta \ln k \le 0.1$ (for $\delta n_s$ we have taken $0 \le \delta n_s \le 0.05$). This ambiguity disappears, of course, in the case of detection {(see below).}

\subsubsection{Measuring the  oscillations \label{subsec:monodetect}}

Let us now study how well we can constrain the axion monodromy parameters in the presence of oscillations with amplitude $\delta n_s = 0.01$ and phase  $\varphi = 0$. The choice of the fiducial value of $\varphi$ is irrelevant for this analysis. For the fiducial frequency we look at two cases: $\delta\ln k=0.1$ and $\delta\ln k=0.01$. The other fiducial parameters are as for the null-test case above.
The constraints on all parameters are collected in Table~\ref{tbl:monodetect}.

\begin{table*}
\centering
\caption{Constraints on  axion monodromy \label{tbl:monodetect}}
\label{mono0.1}
\begin{tabular}{lllll}
\hline
\hline
&  fiducial $\delta\ln k = 0.1$ &  fiducial $\delta\ln k = 0.1$  &  fiducial $\delta\ln k = 0.01$ &  fiducial $\delta\ln k = 0.01$   \\
    & Planck  & Planck + LSS pess. & Planck & Planck + LSS pess. \\
\hline
$\Omega_bh^2$ & $0.02201^{+0.00011}_{-0.00011}$ & $0.02200^{+0.00009}_{-0.00008}$  & $0.02199^{+0.00013}_{-0.00011}$  & $0.02200^{+0.00008}_{-0.00008}$\\
$\Omega_ch^2$  & $0.1128^{+0.0010}_{-0.0010}$ & $0.11277^{+0.00026}_{-0.00025}$  & $0.1129^{+0.0009}_{-0.0010}$ & $0.11283^{+0.00022}_{-0.00025}$\\
$\theta$ & $1.0462^{+0.0002}_{-0.0002}$ & $1.0462^{+0.0002}_{-0.0002}$ & $1.0462^{+0.0002}_{-0.0002}$ & $1.0462^{+0.0002}_{-0.0002}$\\
$\tau_{\rm re}$  & $0.090^{+0.004}_{-0.004}$ & $0.090^{+0.003}_{-0.003}$ & $0.090^{+0.004}_{-0.004}$  & $0.090^{+0.003}_{-0.003}$ \\
$n_s$  & $0.975^{+0.003}_{-0.003}$ & $0.975^{+0.002}_{-0.002}$ & $0.975^{+0.003}_{-0.003}$& $0.975^{+0.002}_{-0.002}$ \\
$\delta n_s$ & $0.0094^{+0.0030}_{-0.0033}$  & $0.0098^{+0.0012}_{-0.0012}$& $0.000^{+0.005+0.014}$ & $0.0098^{+0.0016}_{-0.0016}$ \\
$\delta \ln k$  & $0.101^{+0.006}_{-0.005}$ & $0.100^{+0.003}_{-0.003}$  & unconstrained & $0.01000^{+0.00005}_{-0.00005}$ \\
$\ln(10^{10}A_s)$  & $3.0271^{+0.0076}_{-0.0076}$ & $3.027^{+0.006}_{-0.006}$  & $3.028^{+0.009}_{-0.008}$  & $3.027^{+0.005}_{-0.006}$\\
$\varphi$& unconstrained  & $0.2^{+1.4}_{-1.5}$ & unconstrained  & $0.0^{+2.1}_{-2.1}$ \\
\hline
\end{tabular}
\end{table*}

\begin{figure}
\centering
\includegraphics[width=0.45\textwidth]{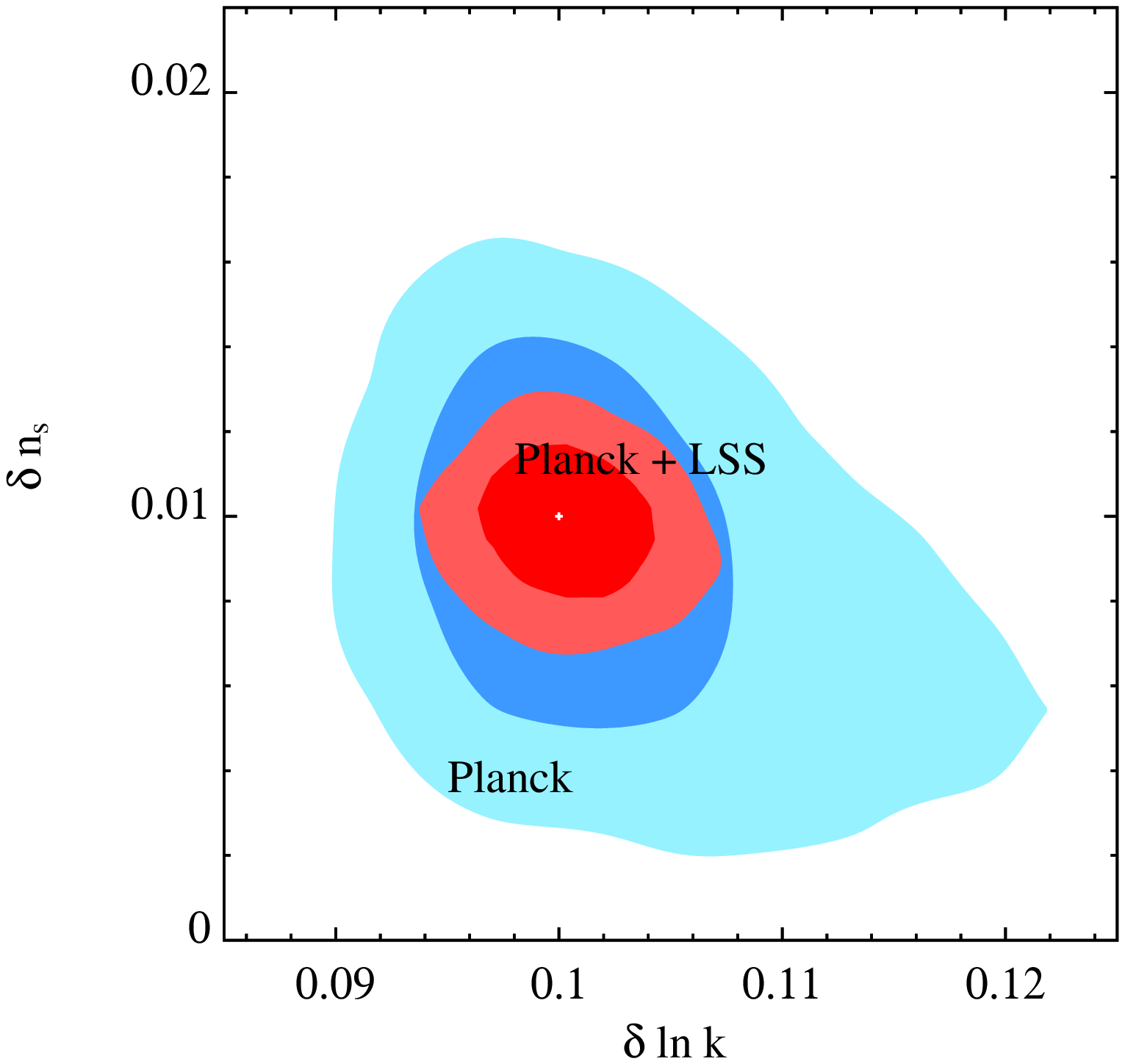}
\includegraphics[width=0.45\textwidth]{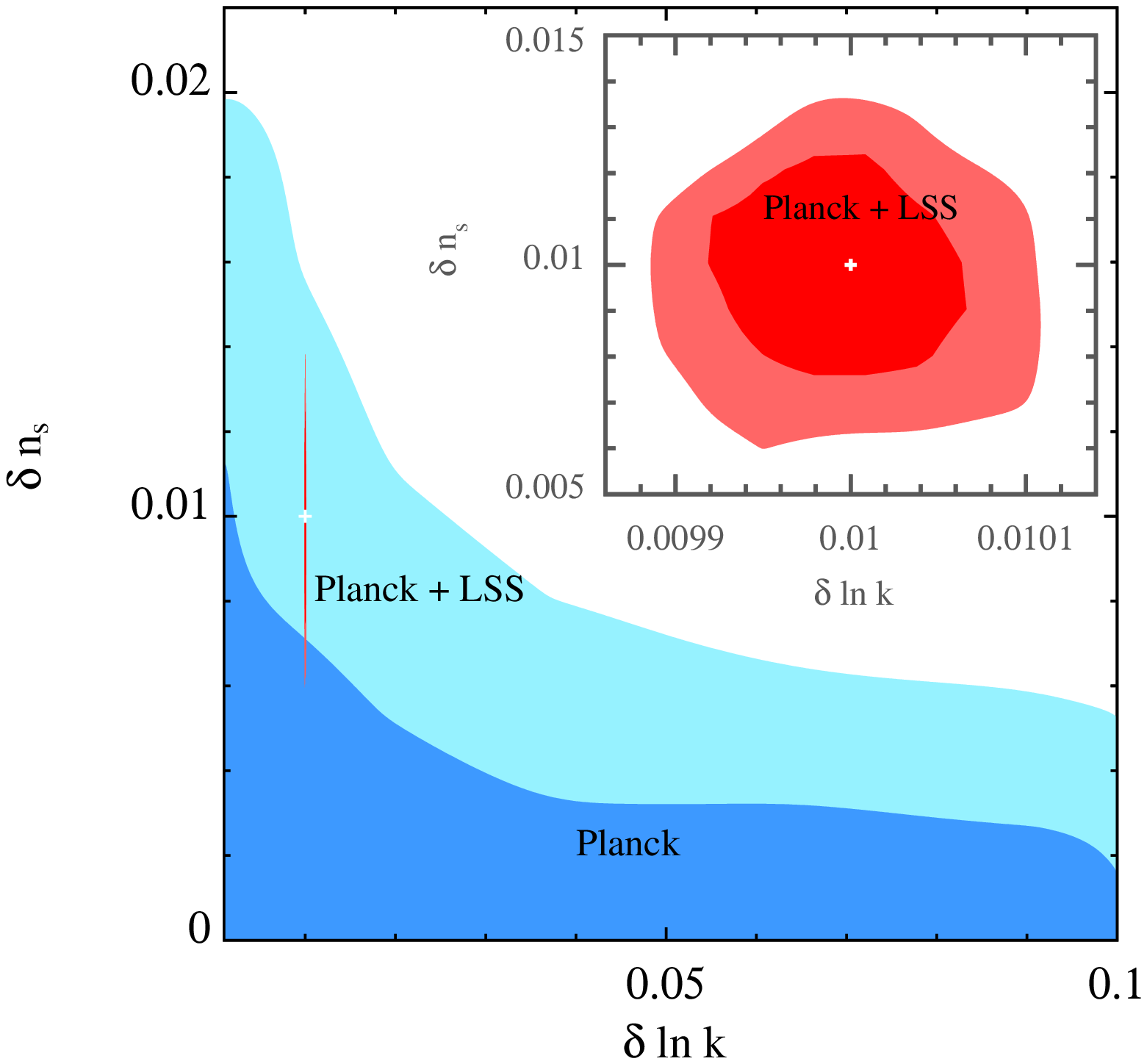}
\caption{The 68.3\% CL and 95.4\% CL constraints on $\delta\ln k$ and $\delta n_s$ with CMB only (blue) or CMB+LSS forecast (red). The white points are the fiducial models: {on the top panel this is $\delta \ln k=0.1$ and $\delta n_s=0.01$; on the bottom panel $\delta \ln k=0.01$ and $\delta n_s=0.01$.} \label{fig:mono}}
\end{figure}

\begin{figure}
\centering
\includegraphics[width=\halffigurewidth]{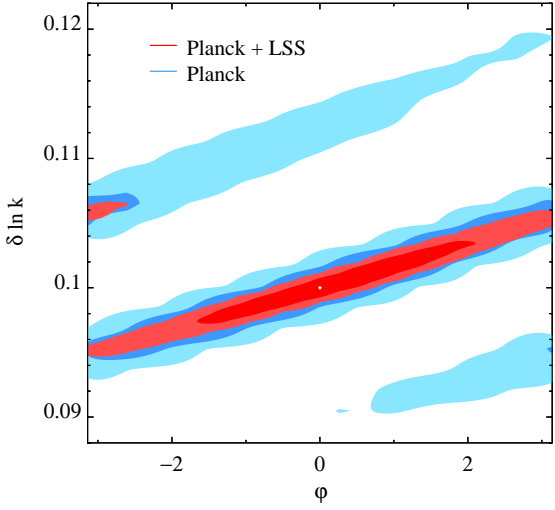}
\caption{The forecasted 68.3\% CL and 95.4\% CL constraints on the ($\varphi$-$\delta \ln k$) plane for Planck-only (blue) and Planck plus a LSS survey (red). We considered fiducial values of $\varphi =0$ and $\delta\ln k= 0.1$.\label{fig:mono0.1phase}}
\end{figure}

For $\delta\ln k=0.1$, Planck-only mock data give a detection of oscillations, which is what expected from the naive estimate, eq.~\eqref{CMB_const}. The inclusion of LSS data can significantly improve the constraints on $\delta \ln k$ and $\delta n_s$, as  shown in Fig.~\ref{fig:mono}. 
Note that, 
in the case of a detection, the inclusion and marginalization 
over the phase parameter $\varphi$ is more important than in the null-test case.
Indeed, as shown   in Fig.~\ref{fig:mono0.1phase} there is  a strong correlation between $\delta\ln k$ and $\varphi$, which is absent in  the null-test case. The usual way of updating the proposal matrix using the covariance matrix, e.g., the method used in CosmoMC,  fails due to the multiple branches of posterior shown in  Fig.~\ref{fig:mono0.1phase}. To solve the problem, we have written a new MCMC engine that allows parameters to have periodic boundary conditions. The MCMC engine is combined with the new $C_\ell$ integrator described in the appendix {\cite{Huang:2012}}.

\begin{figure}
\centering
\includegraphics[width=0.5\textwidth]{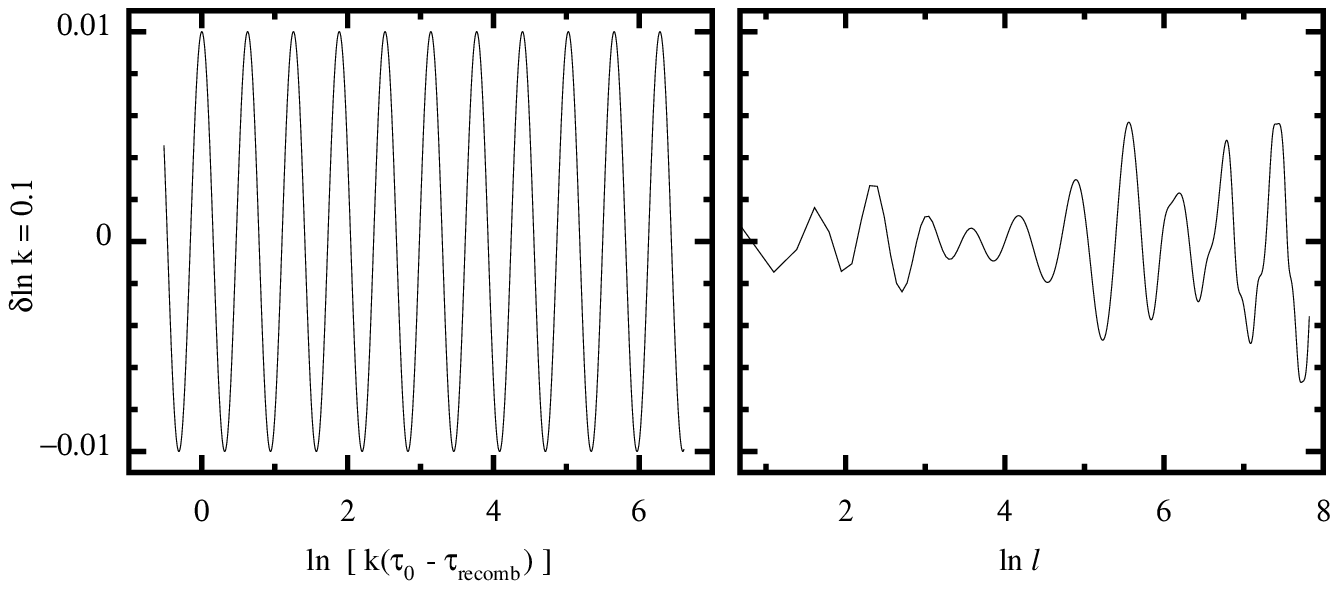}
\includegraphics[width=0.5\textwidth]{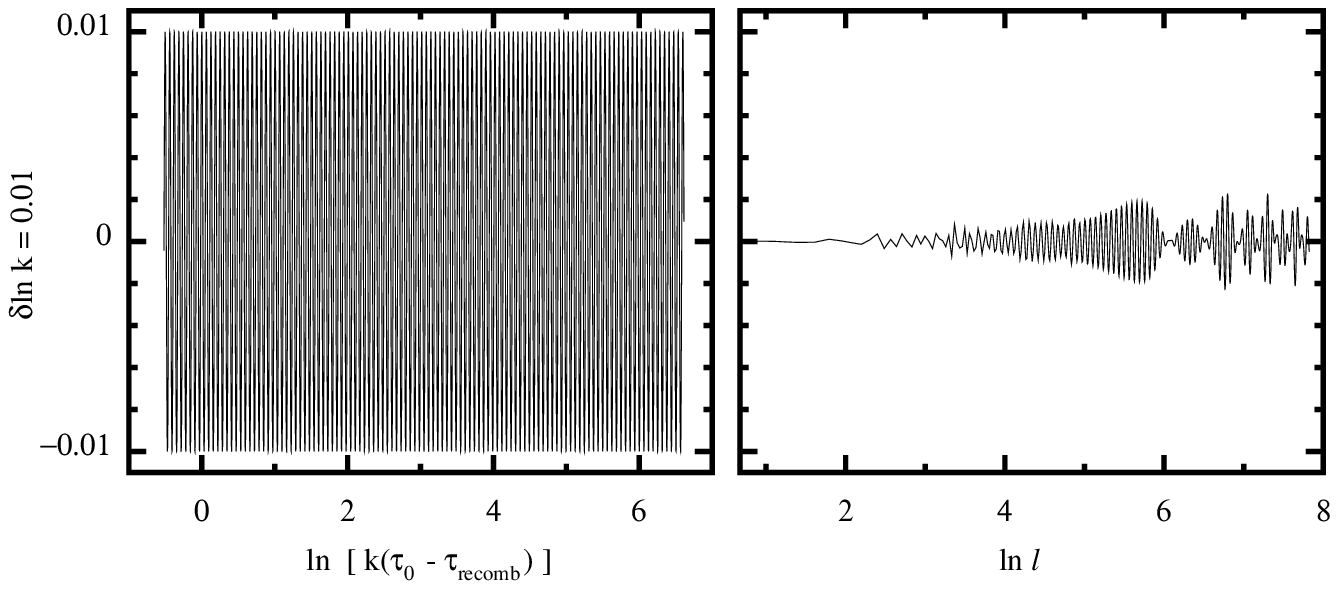}
\includegraphics[width=0.5\textwidth]{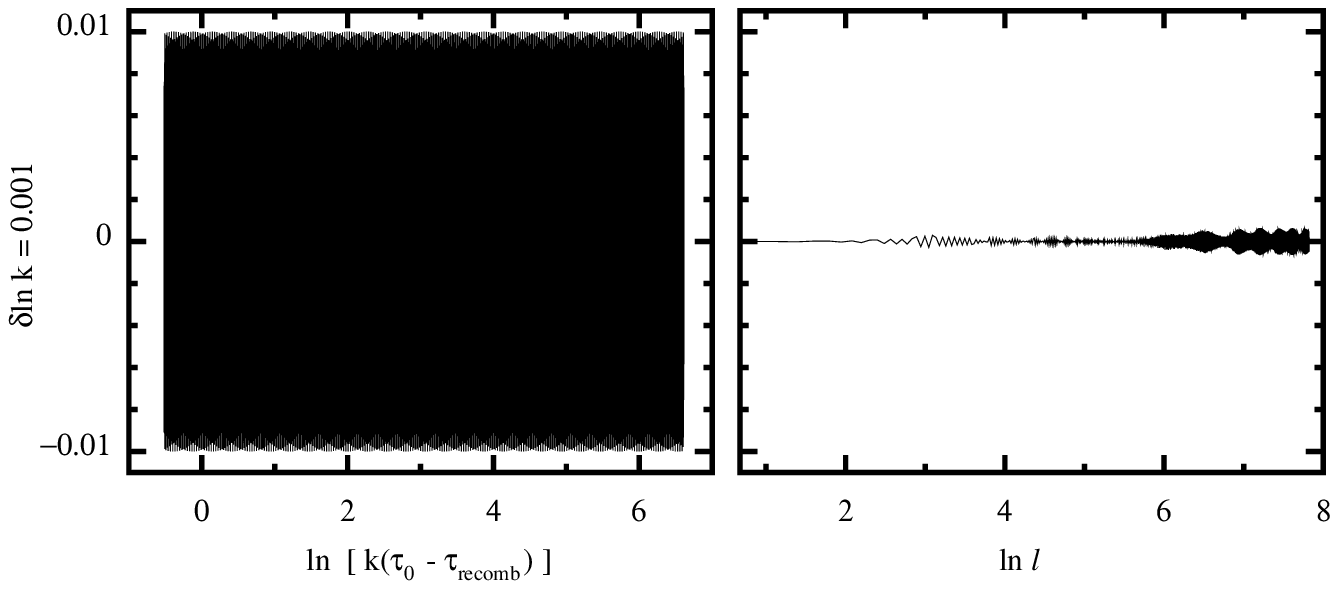}
\caption{The differences in $\ln \primsca$ (left panels) or $\ln C^{TT}_\ell$ (right panels) between a fiducial axion monodromy model with  $\ln \left(10^{10}A_s\right)=3.027$, $n_s = 0.975$, amplitude of cosine modulation $\delta n_s = 0.01$, phase $\varphi=0$  and a smooth power-law spectrum with the same  $A_s$ and $n_s$. From top to bottom, $\delta\ln k = 0.1$, $0.01$, $0.001$, respectively.  \label{fig:pkcls}}
\end{figure}
Let us turn to another choice of fiducial value for the  frequency, $\delta \ln k=0.01$. As shown in the bottom panel of Fig.~\ref{fig:mono}, in this case Planck data cannot measure the oscillations.  This is due to the suppression of features in the $C_\ell$ spectrum occurring in the projection from $k$ to $\ell$ space. 
To explicitly see this suppression, we compute the CMB temperature-temperature angular power spectrum $\ln C^{TT}_\ell$   {for a primordial power spectrum $\ln \primsca$} with and without monodromy sinusoidal modulation. The differences between the two, for $\delta\ln k = 0.1$, $0.01$ and $0.001$, are shown in Fig.~\ref{fig:pkcls}. Apart from the modulation due to the acoustic transfer, which cannot be capture by the simple analytic argument of Sec.~\ref{sec:CMB} \cite{Adshead/etal:2011b}, the suppression due to the projection from $k$ to $\ell$ space shown on the right-hand panels is roughly given by $\sqrt{\delta \ln k}$, which confirms our discussion of Sec.~\ref{sec:CMB}. Moreover, at low $\ell$, $\ell \lesssim (\delta \ln k)^{-1}$, the oscillations in $k$-space disappear in $\ell$-space: indeed,  because of the discreteness of $\ell$ they simply cannot be sampled.

Note that the error bar on the parameter $\delta\ln k$ is extremely tiny in the case of detection from CMB + LSS shown in the lower panel of Fig.~\ref{fig:mono}. One can estimate this error as 
\begin{equation}
  \sigma_{\delta\ln k} \sim  N_{\rm osc}^{-1} \, \delta\ln k \, , \label{eq:sigmadlnk}
\end{equation}
where $N_{\rm osc}$ is the number of oscillations in the observable window in $\ln k$. This estimate can be found by considering the shift in $\delta\ln k$ inducing a change in $N_{\rm osc}$ of order unity. Indeed, a much larger change of $N_{\rm osc}$ makes most of the oscillations out of phase, hindering the detection. Thus, using that $N_{\rm osc} \propto (\delta\ln k)^{-1}$ we find eq.~\eqref{eq:sigmadlnk}. Both panels of Fig.~\ref{fig:mono} confirm this estimate.

Note also  that from eq.~\eqref{CMB_const} we would expect a detection. However, this estimate is too naive to be valid also in the regime where this equation is only marginally satisfied, and one has to rely on the numerical analysis. Adding the LSS improves considerably the figure. Indeed, in this case we can detect oscillations and  obtain rather tight constraints on both $\delta\ln k$ and $\delta n_s$.

\begin{figure}
\includegraphics[width=0.5\textwidth]{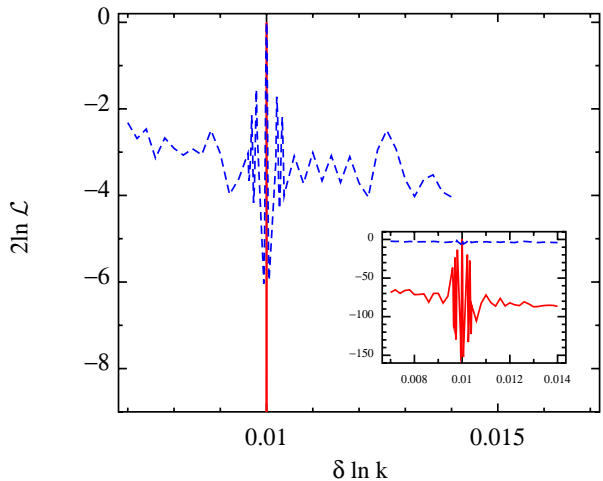}
\caption{Slices of the CMB likelihood (dashed blue lines) and the large-scale structure likelihood (solid red lines) for varying $\delta \ln k$. The other parameters are all fixed at the fiducial values. The fiducial $\delta\ln k=0.01$ is where the spike of likelihood presents. Both likelihoods are normalized to 1 at their maxima. The main frame shows the large-likelihood part that includes the CMB likelihood slice and a sharp spike of the large-scale structure  one, while the lower-right panel zooms out the global structure of the large-scale structure likelihood slice.\label{fig:slicelike}}
\end{figure}

If we vary $\delta\ln k$ and keep all the other parameters fixed at their fiducial values, the form of the likelihood as a function of $\delta \ln k$ is very different from a Gaussian: it is typically sharply spiked around the fiducial value of $\delta\ln k$ and remains almost constant  far from the fiducial value of $\delta\ln k$.  An example is shown in Fig.~\ref{fig:slicelike}. This form of likelihood cannot be studied by standard Fisher matrix analysis. Indeed, in this case searching the maximum of the likelihood is more similar to the process of tuning a radio channel to search for the correct frequency.
Since a tiny change of $\delta\ln k$ can destroy the tuned-in resonance between the data and the theoretical prediction, the error bar on $\delta\ln k$ is either tiny in the case of a detection, or very large in the case of no detection. 

This is exactly what happens in the case of fiducial $\delta \ln k=0.01$ discussed above. In the Planck-only case the 
parameter space is dominated by the tuned-out situation, where the frequency and phase of the oscillations due to monodromy in $ \hat C_\ell^{XY}$ and $C_\ell^{XY}$ of eq.~\eqref{chisq_CMB} are different. In this case the $\chi^2$ of \eqref{chisq_CMB} is minimized by models with small $\delta n_s$ ($\chi^2$  will be roughly proportional to $(\delta \hat n_s)^2 + (\delta n_s)^2$, where the cross term vanishes due to uncorrelated phases). 
Thus, after marginalization over all the other parameters the fiducial value $\delta n_s=0.01$ is not a better fit than $\delta n_s=0$. Indeed,  as shown in Fig.~\ref{fig:mono}, the fiducial model is not even in the marginalized 1-$\sigma$ contour.

\section{Conclusion \label{sec:conclusion}}

In this paper we studied how   planned  stage IV galaxy  redshift surveys (according to the classification of \cite{Albrecht/etal:2006}) with characteristics not too dissimilar from those of, e.g., the  Euclid mission,  will improve the constraints on inflationary models that one obtains from CMB data alone,
in various situations. For slow-roll models and in the more general case of a smooth inflaton potential over the field range of a Planck mass (see Sec.~\ref{sec:smooth}), we have studied the forecasted constrains on the spectral index and its running. From the expected signal-to-noise of the Planck satellite data we have found that by adding  LSS galaxy survey data  we obtained an improvement of about a factor of two, in broad agreement with \cite{Takada/etal:2006}.

Then, in Sec.~\ref{sec:glitches} we have considered models predicting sharp features in the power spectrum. In this case galaxy surveys will be crucial to detect and measure features. Indeed, what we measure with the CMB is the angular power spectrum of the anisotropies in the 2-D multipole space, which is a projection of the power spectrum in the 3-D  momentum space. As explained in Sec.~\ref{sec:glitches},  features at large $\ell$'s  and for small width in momentum space get smoothed during this projection {\cite{Adshead/etal:2011}}. Thus, future galaxy surveys will be able to measure features that are invisible to CMB observations. The main limitation on the width of features measured  using LSS  comes from the size of the volume of the survey, the smallest detectable feature being of the order of the inverse cubic root of this volume. The other limitation comes from the number of modes contained in this volume. In order to improve  measurements of oscillations, it will be important to have a better control on the maximum wavemode that can be used in the analysis. This can be done only by reaching a better understanding on redshift distortions,  non-linear physics at small scales and galaxy bias. These findings are robust to  small changes of the survey characteristic especially for models which give a smooth power spectrum. Models with  sharp features or oscillations however are very sensitive to the fraction of the observed sky (i.e.~to the survey volume) as it is clear from Sec.~\ref{sec:LSS_an}.

For models predicting oscillations in the power spectrum we have found that future LSS will push the amplitude of observability down to $\delta n_s \simeq 10^{-3}$, provided that the frequency of oscillations is $0.01\lesssim \delta\ln k \lesssim 1$. (For $\delta\ln k\gg 1$ the oscillations again become unmeasurable due to the limited range of observable scales.) Quantitatively similar conclusions have also been reached earlier in \cite{Hamann/etal:2008}.

Note that hints of monodromy modulation in WMAP and ACT CMB power spectrum has been recently reported in Refs.~\cite{Aich/etal:2011, Meerburg/etal:2011}. Since the best-fit $\delta\ln k$ claimed in Ref.~\cite{Aich/etal:2011} is of order $10^{-2}$, caution needs to be taken for the accuracy of $C_\ell$ integration, the marginalization and the MCMC convergency. Most notably, a significant improvement of $\chi^2$ at some best-fit point does {\it not} necessarily indicate a detection of oscillations, as the likelihood cannot be approximated as a multivariate Gaussian. Indeed, the absolute height of the likelihood spike alone cannot determine whether there is a detection or not. What matters is the product of the height of the likelihood spike and the volume that it occupies -- see Fig.~\ref{fig:slicelike} and the discussion in the last few paragraphs in subsection~\ref{subsec:monodetect}. 

For this analysis one of us (ZH) has developed a $C_\ell$ integrator,  a modified MCMC engine that allows parameters to have periodic boundary conditions and a forecast mock data generator that does the calculation described in Section~\ref{sec:nsnrun}.  We briefly review their main features in the appendix. The package with all  these codes is a self-contained tool to forecast  constraints using large-scale structure, CMB and supernovae  -- although the latter have not been used in this paper. It is publicly released at \url{http://www.cita.utoronto.ca/~zqhuang/CosmoLib} and described in more details in an accompanying paper \cite{Huang:2012}.

\subsection*{Acknowledgments:} We wish to thank Xingang Chen, Guido D'Amico, Lam Hui, Liam McAllister, Cristiano Porciani and Emiliano Sefusatti for useful advice and discussions.  We also thank Henk Hoekstra, Tom Kitching and  Will Percival for  a careful reading of the draft.

\appendix

\section{The Numerical Package \label{sec:tech}} 

The modified MCMC engine and the new integrator to compute more accurately the CMB angular power spectra for models with rapid oscillations is presented in detail in Ref.~\cite{Huang:2012}. Here we summarize  its main features.

For the MCMC engine, to achieve quicker convergence in the case that the likelihood is a periodic function of some parameter $\varphi$, in each random-walk step we have mapped the proposed new value of $\varphi$ into $[-\pi, \pi)$ by adding integer number of $2\pi$'s to $\varphi$, instead of using the usual rule that $\varphi$ is rejected when it exceeds the boundaries.

\begin{figure}
  \includegraphics[width=0.5\textwidth]{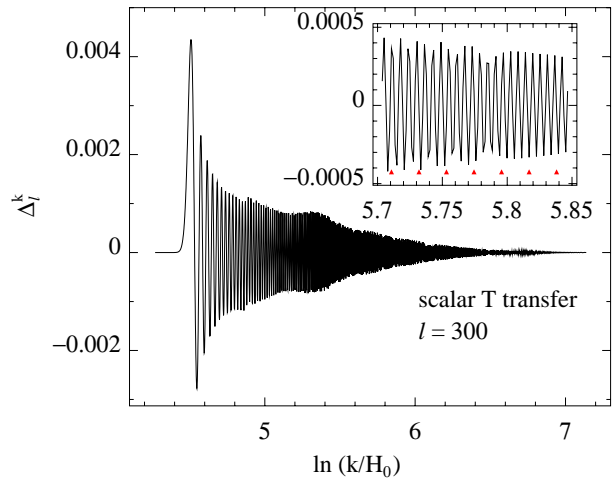}
  \caption{The temperature transfer function  $\Delta^k_\ell$ for a fixed $\ell = 300$ \label{fig:trans}}
\end{figure}

The challenging task is to accurately compute the angular CMB spectrum for each multipole $\ell$, in the presence of oscillations in the primordial power spectrum $\primsca(k)$. The angular spectrum for the temperature anisotropies is given by 
\begin{equation}
C_\ell = \int |\Delta_\ell^k |^2 \primsca(k) d\ln k \;,  \label{eq:clint}
\end{equation}
where $\Delta_\ell^k$ is the temperature transfer function, which is typically oscillatory. Since $\Delta_\ell^k$ is evaluated numerically, its sampling is time consuming. Indeed, in modern fast CMB codes -- such as CAMB \cite{Lewis/etal:2000}, CLASS \cite{Lesgourgues:2011, Blas/etal:2011}, or CMBEasy \cite{Doran:2005} -- the integral above is computed by sampling $\Delta_\ell^k$ using a step size which can be typically much larger than the oscillation period. For instance, in Fig.~\ref{fig:trans} we show an example of $\Delta_\ell^k$ for a fixed $\ell = 300$. A typical sampling scheme is shown by the red solid triangles in the upper-right panel, which zooms-in part of the figure. If $\primsca (\ln k)$ is a smooth function -- i.e.~a low-pass window -- sparse sampling of $\Delta_\ell^k$ is enough. 

However, when $\primsca(k)$ has local sharp features, at least a few samples of $\Delta_\ell^k$ are needed per $\delta\ln k$, the typical width of the features. For instance, if our goal is to sample features with width $\delta \ln k \sim 10^{-3}$, the required sampling frequency is typically $\sim 20 $ to  $100$ times higher than that used for a smooth $\primsca(k)$. Furthermore, as we wish to compute the $C_\ell$'s  for each $\ell$ rather than interpolating it over few tens of $\ell$'s, its computation is typically  few thousands times more expensive than in the smooth-$\primsca(k)$ case. A final complication is due to the fact that, if all the transfer functions and the precomputed $j_\ell(x)$ tables are to be stored,  one has also to face a memory barrier that cannot be easily bypassed. 
For these reasons, simply increasing the $\ell$ and $k$ resolution in standard codes such as CAMB, CLASS or CMBEasy, will not be  efficient enough for the purpose of scanning the whole parameter space. 

Here we  compute the $C_\ell$'s in equation~(\ref{eq:clint}) $\ell$ by $\ell$. We chose the sampling period in $\ln k$ to be at least a few times smaller than $\delta \ln k$. To speed up the computation and avoid memory problems, our strategy is the following.  
We first compute two neighboring $C_\ell$'s by brute force (see below). Two arrays of spherical Bessel functions $j_{\ell+1}[k(\tau_0 -\tau)]$ and $j_\ell[k(\tau_0-\tau)]$ are stored in the memory for each $(k, \tau)$ sample used in the computation. Then we compute $C_{\ell-1}$. To do that, we directly compute $j_{\ell -1}$ using the recurrence relation 
\begin{equation}
  j_{\ell -1}(x) = \frac{2\ell +1}{x} j_\ell(x) - j_{\ell +1} (x) \ .\label{eq:jlrecur}
\end{equation}
The values of $j_{\ell +1}$ are then discarded to free the memory. Again, using $j_\ell$ and $j_{\ell-1}$ we then calculate $j_{\ell -2}$ and hence $C_{\ell -2}$. This downward iteration is very stable for a few tens of steps, after which we need to recompute another couple of neighboring $C_\ell$'s and iterate downward  again. 

The initial neighboring $j_{\ell}$'s are calculated using precomputed 25-th order Chebyshev fitting formulas. (For the rapidly oscillating part at $x\gg l$, the phase and amplitude of oscillations are fitted using Chebyshev polynomials.) Chebyshev fitting is slightly slower than the cubic-spline fitting used in other public CMB codes, but it is more memory-efficient and more accurate -- it has an accuracy $\sim 10^{-8}$ -- and allows more downward iterative steps. Finally, note that the algorithm proposed here is more efficient both CPU-wise and memory-wise, enhancing the speed of of $\ell$-by-$\ell$ computation of $C_\ell$'s by a factor of $\sim 10$ to $30$.

\end{document}